\documentclass[aps,prl,twocolumn,superscriptaddress, nofootinbib]{revtex4-2}
\usepackage{graphicx}
\usepackage{latexsym}
\usepackage{amsmath}
\usepackage{upgreek}
\usepackage{bibunits}
\defaultbibliography{quant_frustr}
\usepackage{subfigure}
\usepackage{MnSymbol}
\usepackage{lineno}
\usepackage{bm}
\usepackage{bbold}
\usepackage[unicode=true,
 bookmarks=false,
 breaklinks=false,pdfborder={0 0 1},backref=false,colorlinks=true]
 {hyperref}
\hypersetup{
 linkcolor=[rgb]{0,0,1},citecolor=[rgb]{0,0,1},urlcolor=[rgb]{0,0,1}}
\usepackage{multirow}
\usepackage{color}
\usepackage{comment}
\usepackage{xcolor}
\usepackage{soul}
\usepackage{tikz}

\usepackage{braket}

\begin{document}

\title{Quantifying hole-motion-induced frustration in doped antiferromagnets by Hamiltonian reconstruction}

\author{Henning Schl\"omer}
\email{H.Schloemer@physik.uni-muenchen.de}
\affiliation{Department of Physics and Arnold Sommerfeld Center for Theoretical Physics (ASC), Ludwig-Maximilians-Universit\"at M\"unchen, M\"unchen D-80333, Germany}
\affiliation{Munich Center for Quantum Science and Technology (MCQST), D-80799 M\"unchen, Germany}
\affiliation{ITAMP, Harvard-Smithsonian Center for Astrophysics, Cambridge, MA, USA}
\author{Timon A. Hilker}
\affiliation{Munich Center for Quantum Science and Technology (MCQST), D-80799 M\"unchen, Germany}
\affiliation{Max-Planck-Institut f\"ur Quantenoptik, D-85748 Garching, Germany}
\author{Immanuel Bloch}
\affiliation{Department of Physics and Arnold Sommerfeld Center for Theoretical Physics (ASC), Ludwig-Maximilians-Universit\"at M\"unchen, M\"unchen D-80333, Germany}
\affiliation{Munich Center for Quantum Science and Technology (MCQST), D-80799 M\"unchen, Germany}
\affiliation{Max-Planck-Institut f\"ur Quantenoptik, D-85748 Garching, Germany}
\author{Ulrich Schollw\"ock}
\affiliation{Department of Physics and Arnold Sommerfeld Center for Theoretical Physics (ASC), Ludwig-Maximilians-Universit\"at M\"unchen, M\"unchen D-80333, Germany}
\affiliation{Munich Center for Quantum Science and Technology (MCQST), D-80799 M\"unchen, Germany}
\author{Fabian Grusdt}
\affiliation{Department of Physics and Arnold Sommerfeld Center for Theoretical Physics (ASC), Ludwig-Maximilians-Universit\"at M\"unchen, M\"unchen D-80333, Germany}
\affiliation{Munich Center for Quantum Science and Technology (MCQST), D-80799 M\"unchen, Germany}
\author{Annabelle Bohrdt}
\email{Annabelle.Bohrdt@physik.uni-regensburg.de}
\affiliation{ITAMP, Harvard-Smithsonian Center for Astrophysics, Cambridge, MA, USA}
\affiliation{Department of Physics, Harvard University, Cambridge, Massachusetts 02138, USA}

\date{\today}

\begin{abstract}
{\bf Abstract.} Unveiling the microscopic origins of quantum phases dominated by the interplay of spin and motional degrees of freedom constitutes one of the central challenges in strongly correlated many-body physics. When holes move through an antiferromagnetic spin background, they displace the positions of spins, which induces effective frustration in the magnetic environment.
However, a concrete characterization of this effect in a quantum many-body system is still an unsolved problem. Here we present a Hamiltonian reconstruction scheme that allows for a precise quantification of hole-motion-induced frustration. We access non-local correlation functions through projective measurements of the many-body state, from which effective spin-Hamiltonians can be recovered after detaching the magnetic background from dominant charge fluctuations. The scheme is applied to systems of mixed dimensionality, where holes are restricted to move in one dimension, but $\mathrm{SU}(2)$ superexchange is two-dimensional. We demonstrate that hole motion drives the spin background into a highly frustrated regime, which can quantitatively be described by an effective $J_1-J_2-$type spin model. We exemplify the applicability of the reconstruction scheme to ultracold atom experiments by recovering effective spin-Hamiltonians of experimentally obtained 1D Fermi-Hubbard snapshots. Our method can be generalized to fully 2D systems, enabling promising microscopic perspectives on the doped Hubbard model.
\end{abstract}

\maketitle
{\bf Introduction} 

Microscopically understanding the motion of mobile charge carriers doped into Mott insulators constitutes one of the key open problems in strongly correlated many-body physics. When hopping through an insulating spin environment, holes displace spins along their way, which effectively frustrates the magnetic background. The arising competition of kinetic energy gain via delocalization and associated magnetic energy cost leads to the formation of a plethora of strongly interacting many-body phases~\cite{Lee2006, Keimer2015}, many of which yet seek to be explained on a microscopic footing. The Fermi-Hubbard (FH) model, believed to capture the essential physics of strongly correlated materials, has been subject to intense numerical studies that can resolve the intricate competition between various orders~\cite{LeBlanc2015, Zheng2017, Jiang2019_science, Qin2020, MultiMess2021, Jiang_Kivelson, Arovas2022_rev}. Nevertheless, despite ongoing theoretical and experimental efforts over the past decades, a precise microscopic understanding of the interplay between motional and spin degrees of freedom is still an unsolved task, whose long-sought understanding may help to reveal the origin of high-temperature superconductivity and possibly lead to the discovery of novel pairing mechanisms~\cite{bohrdt2021strong, Hirthe2022}.

Analog quantum simulation, e.g. via ultracold atoms, can shed new light on the microscopic mechanisms underlying strongly correlated quantum many-body states~\cite{Bloch2008, Bloch2012, Cheuk2015, Parsons2015, Gross2017, Schafer2020} and paradigmatic Hamiltonians like the FH model can now be experimentally explored~\cite{Esslinger2010, Hart2015, Cocchi2016, Chiu2019, Hilker2017, Koepsell_science, Bohrdt2020, Hirthe2022}. In particular, these setups allow to perform genuine quantum projective measurements and sample snapshots of the many-body state in the Fock basis, which in turn allow for insights into the wave function beyond averages and local observables. This capability has already been used to unveil highly non-local order parameters and hidden correlations in many-body systems~\cite{Hilker2017, Endres_nonlocal}. 

As we demonstrate in our work, the huge amount of information stored in snapshots of many-body states can further be utilized to disentangle spin and charge sectors through non-local correlation functions, which allow us to recover emergent effective spin-Hamiltonians for parts of the system. The problem of reconstructing a Hamiltonian from measured correlations via machine learning schemes~\cite{DiFranco2009,Zhang2014,Qi2019, Cao2020, anshu2021} has attracted considerable interest in recent years, including certifying quantum simulation devices~\cite{Bairey2019}.

In this article, we present a snapshot-based Hamiltonian reconstruction scheme for the spin channel alone, which removes dominant charge fluctuations~\cite{Hilker2017, Kruis2004} from individual snapshots. This allows us to quantify the effective spin-Hamiltonian, which includes the back-action of mobile dopants on the spin environment. We exemplify the proposed method by considering a system in mixed-dimensions (mixD), where hole motion is restricted to one dimension (1D), but $\mathrm{SU}(2)$ spin-superexchange is two-dimensional (2D). We find that hole hopping drives and stabilizes the spins in a highly frustrated regime, which we show to be accurately described by a $J_1-J_2-$type spin-Hamiltonian. 

Our method is directly applicable to experimental data obtained from ultracold quantum gas microscopes. We showcase this by reconstructing effective Hamiltonians from 1D measurements of the FH model~\cite{Hilker2017}, where spin-charge separation governs the physics of the chains. Furthermore, our insights could be used to effectively simulate the highly frustrated $J_1-J_2$ model in ultracold atom experiments by implementing the mixD setting and post-processing the measurements. 

Our work sheds light on the long-standing question about the interplay of spin- and motional degrees of freedom in strongly correlated materials, and paves the way to gain deep microscopic insights into prototypical systems such as the 2D FH and $t-J$ model. \\

\textbf{Results} 

{\bf The model.} We consider the $t-J$ model in mixD~\cite{Grusdt_tJ, Grusdt_tJz, Schloemer2022, bohrdt2021strong}, described by the Hamiltonian
\begin{equation}
\begin{aligned}
    \hat{\mathcal{H}} =-t \sum_{\braket{\mathbf{i}, \mathbf{j}}_x, \sigma} \hat{\mathcal{P}}_{GW} \big(\hat{c}_{\mathbf{i}, \sigma}^{\dagger} & \hat{c}_{\mathbf{j}, \sigma} + \text{h.c.} \big)\hat{\mathcal{P}}_{GW} + \\ &J \sum_{\braket{\mathbf{i}, \mathbf{j}}} \left( \hat{\mathbf{S}}_{\mathbf{i}} \cdot \hat{\mathbf{S}}_{\mathbf{j}} - \frac{\hat{n}_{\mathbf{i}}\hat{n}_{\mathbf{j}}}{4} \right).
\end{aligned}
\label{eq:mixD_tJ_H}
\end{equation}
Here, $\hat{c}_{\mathbf{i}}^{(\dagger)}$, $\hat{n}_{\mathbf{i}}$ and $\hat{\mathbf{S}}_{\mathbf{i}}$ are fermionic annihilation (creation), charge density, and spin operators on site $\mathbf{i}$, respectively; $\braket{\mathbf{i}, \mathbf{j}}_{(x)}$ denotes a nearest-neighbor (NN) pair on a 2D square lattice (with subscript $x$ indicating a NN pair only along the $x$-direction), and $\hat{\mathcal{P}}_{GW}$ is the Gutzwiller operator projecting out states with double occupancy. 
The mixD setting, Eq.~\eqref{eq:mixD_tJ_H}, has successfully been implemented in ultracold atom setups using strong tilted potential gradients~\cite{Hirthe2022}, which effectively restrict hole motion perpendicular to the gradient direction while spin-spin interactions remain 2D~\cite{Trotzky, Dimitrova}.

Recently, we demonstrated how hidden AFM correlations in the mixD $t-J$ model result in the formation of a remarkably resilient stripe phase (i.e. a coupled charge- and spin-density wave~\cite{White_stripes, TRANQUADA20121771}), with critical temperatures on the order of the magnetic coupling $J$~\cite{Schloemer2022}. Above these critical temperatures of charge- and spin-density wave formation, holes were found to form a deconfined chargon gas, i.e., a phase without order~\cite{Grusdt_tJz, Schloemer2022}.

\begin{figure}
\centering
\includegraphics[width=0.87\columnwidth]{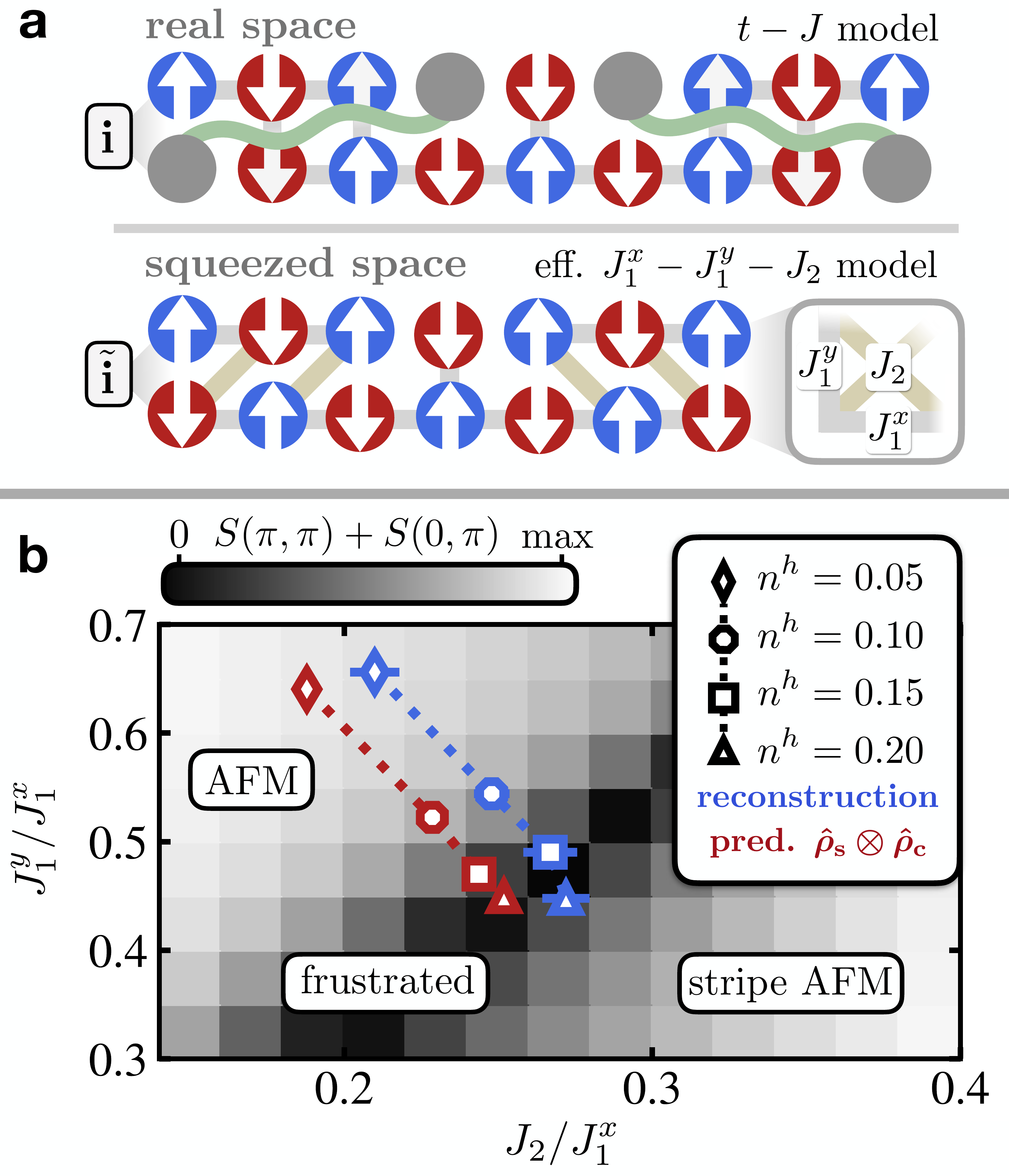}
\caption{\textbf{Hole-motion-induced spin frustration}. \textbf{a}, Schematic of how hole hopping induces frustration in the spin background. Upper panel: Snapshot of holes moving through a N\'eel background. Spatial separation of holes on neighboring legs lead to the formation of geometric strings (green wiggly lines) costing magnetic energy. Lower panel: Upon transforming the snapshot to squeezed space, originally vertical bonds $J_1^y$ in between two holes on neighboring legs become effective diagonal couplings $J_2$. The resulting energy penalty of aligned diagonal spins leads to frustration in the magnetic background. \textbf{b}, Hamiltonian reconstruction results (blue) for hole dopings $n^h = 0.05 \dots 0.2$ of a mixD $t-J$ ladder with $t/J = 3$, $L_x \times L_y = 20 \times 2$ and $T/J = 5/3 \sim 1.67$. Reconstructions of squeezed space according to the input $J_1^x-J_1^y-J_2$ Heisenberg Hamiltonian, Eq.~\eqref{eq:J1J2}, are presented in a $J_1^y/J_1^x, J_2/J_1^x$ diagram. Light regions in the background signal the presence of either AFM or stripe AFM order in the purely magnetic $J_1^x-J_1^y-J_2$ model in the ground state by plotting the sum of the spin structure factors $S(\pi,\pi) + S(0,\pi)$. Dark regions correspond to a highly frustrated regime without apparent order. Upon doping the system, the background spins are driven into a strongly frustrated state. Error bars correspond to the standard error to the mean when averaging over ten reconstruction runs. Red connected symbols show theoretical expectations assuming no spin-hole correlations in the mixD $t-J$ model, i.e., $\hat{\rho} = \hat{\rho}_{\text{s}} \otimes \hat{\rho}_{\text{c}}$, evaluated via Eq.~\eqref{eq:pred_J1x} and~\eqref{eq:pred_Jn}.}
\label{fig:recon}
\end{figure}

In the following, we focus on the latter regime, and study how hole motion distorts the spins in the background. The effect is qualitatively depicted in Fig.~\ref{fig:recon}~\textbf{a}. The upper panel shows an (idealized) real space snapshot of holes moving through an AFM N\'eel background. Bonds correspond to AFM interactions in the instantaneous charge configuration, illustrated by gray lines. In between holes on neighboring legs, spins are aligned, leading to a linearly increasing magnetic energy penalty via the formation of geometric strings~\cite{GrusdtX, Grusdt_strings, Chiu2019, Bohrdt_ML} (depicted by green wiggly lines). 

As a direct consequence of the restricted charge motion to 1D, spins can be relabeled by the new positions they have after moving all holes to the right in each chain -- resulting in a distinct definition of squeezed space~\cite{Kruis2004, OgataShiba}. More formally, consider a Fock state $\bigotimes_{y} \ket{\sigma_{[1,y]}, \sigma_{[2,y]}, \dots, \sigma_{[L_x, y]}}$, where $\{0, \uparrow, \downarrow\} \ni \sigma_{x,y}$ is the single particle basis of the $t-J$ model. These local spin charge configurations are relabeled upon squeezing, whereby each Fock state is now given by $\bigotimes_{y} \ket{\tilde{\sigma}_{[\tilde{1},y]}, \sigma_{[\tilde{2},y]}, \dots, \sigma_{[\tilde{L}_x, y]}} \otimes \hat{h}^{\dagger}_{[x_1, y]} \dots \hat{h}^{\dagger}_{[x_{N_y}, y]} \ket{0}$~\cite{Grusdt_tJz}. Here, $\tilde{\sigma}_{[\tilde{x}, y]} = \uparrow, \downarrow$ (but note that $\tilde{\sigma}_{[\tilde{x}, y]} \neq 0$) denotes spins on the squeezed lattice $\tilde{x} = 1, \dots, L_x - N_y$, where $N_y$ is the number of holes in rung $y$, and $\hat{h}_{[x,y]}$ creates a hard core fermionic chargon at site $\mathbf{i} = [x,y]$. By squeezing the spins out, spins on the squeezed and real space lattice relate as $\tilde{\sigma}(\tilde{x}, y) = \sigma(\tilde{x} + \sum_{j<\tilde{x}} n^h_{[j,y]}, y)$, where $n^h_{[x,y]}$ refers to the number of chargons at real space lattice site $[x,y]$. The lower panel of Fig.~\ref{fig:recon}~\textbf{a} illustrates the squeezing process, where the initial N\'eel order is restored in the isolated spin background. However, interactions on diagonal bonds emerge (ocher lines), which cause geometric frustration of the spins in squeezed space. From now on, we refer to lattice sites in real and squeezed space by $\mathbf{i}$ and $\tilde{\mathbf{i}}$, respectively. \\

\begin{figure*}
\centering
\includegraphics[width=0.95\textwidth]{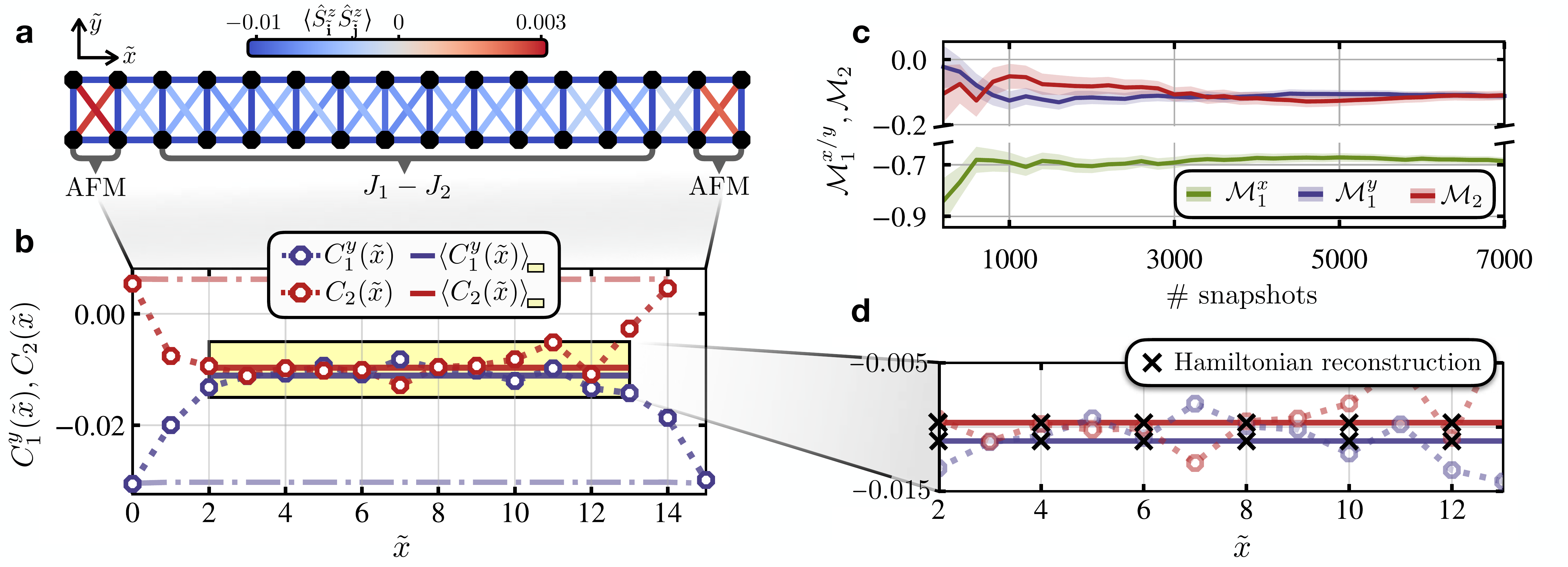}
\caption{\textbf{Correlations in squeezed space \& Hamiltonian reconstruction.} \textbf{a}, Spin-spin correlations $\braket{\hat{S}_{\tilde{\mathbf{i}}}^z \hat{S}_{\tilde{\mathbf{j}}}^z}$ on the squeezed lattice of an original $20\times 2$ mixD $t-J$ ladder, with $n^h = 0.2$, $T/J = 5/3 \sim 1.67$, and using $20,000$ snapshots. In the bulk of squeezed space, hole motion distorts the spin background, leading to negative correlations across diagonals. In this region, effective $J_1-J_2$ physics is expected, as captured by the Hamiltonian Eq.~\eqref{eq:J1J2}. As holes are rarely located at the open boundaries of the system, correlations are left almost undisturbed and are of AFM type. Correlations along nearest neighbors in $x$ go beyond the cutoff of the colorbar. \textbf{b}, Rung $C_1^y(\tilde{x}) = \braket{\hat{S}^z_{[\tilde{x},\tilde{0}]} \hat{S}^z_{[\tilde{x},\tilde{1}]}}$ and diagonal $C_2(\tilde{x}) =  \braket{\hat{S}^z_{[\tilde{x},\tilde{0}]} \hat{S}^z_{[\tilde{x}+1,\tilde{1}]}} + \braket{\hat{S}^z_{[\tilde{x}+1,\tilde{0}]} \hat{S}^z_{[\tilde{x},\tilde{1}]}}$ correlations. In the central bulk region of the ladder, correlations are approximately constant, the average being used as the input for $J_1^x-J_1^y-J_2$ Hamiltonian reconstructions. In particular, we discard the two outer sites in squeezed space, as illustrated by the yellow box. Dashed-dotted lines correspond to rung and diagonal correlations for a nearest-neighbor Heisenberg model with $\beta J_1^x = \beta J_1^y \sim 0.44$, which captures the physics at the edges, but fails to describe the correlations in the bulk of squeezed space. The introduction of diagonal (frustrating) bonds is hence an essential step to describe the spin system on the squeezed lattice. \textbf{c}, Summed correlations in the boxed bulk in \textbf{b} along rungs, legs and diagonals for varying snapshot set sizes. Light regions correspond to the standard error to the mean. After a few thousand measurements, convergence of the correlator proxies is reached. \textbf{d}, Results show perfect agreement between the correlations emerging from a reconstructed effective $J_1^x-J_1^y-J_2$ Heisenberg model (black crosses) and bulk averaged correlations of the doped mixD model in squeezed space (solid lines).}
\label{fig:recon_details}
\end{figure*}

\textbf{Characterizing the spin state in squeezed space.} In order to quantify the arising frustration on the squeezed lattice, we simulate the mixD $t-J$ model, Eq.~\eqref{eq:mixD_tJ_H}, at finite temperature and fixed doping using imaginary time evolution schemes (purification) via matrix product states (MPS)~\cite{Schollwoeck_DMRG, Paeckel_time}. For faster, more controllable numerics and to prevent post-selection of snapshots, we explicitly implement the system's enhanced $\mathrm{U}(1)$ symmetries in each ladder leg, i.e., we work in an ensemble where we allow for thermal spin fluctuations but keep the number of holes in each ladder leg constant~\cite{Schloemer2022}. In particular, we simulate the mixD $t-J$ model at intermediate temperature $T/J= 5/3$ ($\beta J = 0.6$), which lies inside the chargon gas phase (i.e. no charge- and spin-density waves form) and, furthermore, is in a temperature regime accessible for quantum gas microscopes.

From the thermal MPS at inverse temperature $\beta$, we sample uncorrelated snapshots of the corresponding Gibbs state~\cite{Ferris2012, Buser2022}. After post-processing the individual measurements by squeezing out the holes, spin-spin correlations $\braket{\hat{S}_{\tilde{\mathbf{i}}}^z \hat{S}_{\tilde{\mathbf{j}}}^z}$ can directly be evaluated in squeezed space. In Fig.~\ref{fig:recon_details}~\textbf{a}, nearest-neighbor as well as diagonal spin-spin correlations are shown on the squeezed lattice. In the bulk of the squeezed ladder, both nearest neighbor as well as diagonal correlators are negative and comparable in magnitude, signaling strong frustration in the spin background induced by the motion of the holes. In contrast, at both edges of the ladder, low average hole concentrations lead to only marginal perturbations of the spin background -- resulting in AFM type correlations that are negative (positive) along nearest (diagonal) neighbors. 

Fig.~\ref{fig:recon_details}~\textbf{b} shows nearest-neighbor rung (blue) and diagonal (red) correlators. Dashed-dotted lines correspond to rung and diagonal correlations for a Heisenberg ladder with solely nearest-neighbor couplings (where $\beta J_1^x = \beta J_1^y = 0.44$), which describe the physics at the edges qualitatively well, but fail to reproduce the measured correlations in the bulk. 
Due to the frustrating effect of the hopping holes, diagonal couplings need to be taken into account to accurately capture the physics of the squeezed background. To this end, we introduce an effective $J_1^x-J_1^y-J_2$ Heisenberg model~\cite{Starykh2004}, given by the Hamiltonian
\begin{equation}
\hat{\mathcal{H}}_{\{J_H\}} = \sum_{\mu = x,y} J_1^{\mu} \sum_{\braket{\tilde{\mathbf{i}},\tilde{\mathbf{j}}}_{\mu}} \hat{\mathbf{S}}_{\tilde{\mathbf{i}}} \cdot  \hat{\mathbf{S}}_{\tilde{\mathbf{j}}} + J_2 \sum_{\llangle \tilde{\mathbf{i}},\tilde{\mathbf{j}} \rrangle_{\text{diag}}}  \hat{\mathbf{S}}_{\tilde{\mathbf{i}}} \cdot  \hat{\mathbf{S}}_{\tilde{\mathbf{j}}}.
\label{eq:J1J2}
\end{equation}
Here, $\hat{\mathbf{S}}_{\tilde{\mathbf{i}}}$ is the spin-1/2 operator at site $\tilde{\mathbf{i}}$, and $\{J_H\} = \{J_1^{x}, J_1^y, J_2\}$ are the coupling strengths of neighbouring spins in $x,y$ and diagonal direction on the squeezed lattice, respectively. 

To quantitatively pin down the strength of the arising frustration, we perform a Hamiltonian reconstruction with input Hamiltonian $\hat{\mathcal{H}}_{\{J_H\}}$, Eq.~\eqref{eq:J1J2}, together with the measured spin-spin correlations in squeezed space. Couplings in $\hat{\mathcal{H}}_{\{J_H\}}$ are chosen to be homogeneous throughout the bulk of squeezed space, justified by the approximately constant behaviour of correlations in the bulk region of Fig.~\ref{fig:recon_details}~\textbf{b} -- solid lines are averages of the correlations over the marked box. Due to the underlying $\mathrm{SU(2)}$ symmetry of the mixD $t-J$ model, many-body snapshots along a single spin axis -- here chosen along $z$ -- are sufficient to reconstruct the full effective Hamiltonian, Eq.~\eqref{eq:J1J2}. Results of the reconstruction correspond to the parameter configuration $\{J_H\}$ that best describe the measured correlations.

On a more formal footing, we follow the procedure introduced in~\cite{anshu2021} and minimize the objective function $\mathcal{G}$ over all possible coupling parameters $\{J_H\}$,
\begin{equation}
\mathcal{G} = \ln Z(\beta, \{J_H\}) + 3 \beta\left( \sum_{\mu=x,y} J_1^{\mu} \mathcal{M}_1^{\mu} + J_2 \mathcal{M}_{2} \right),
\label{eq:obj}
\end{equation}
with $Z=\text{Tr}[e^{-\beta \hat{\mathcal{H}}_{\{J_H\}}}]$ the partition function and $\mathcal{M}_1^{\mu} =  \sum^{\prime}_{\braket{\tilde{\mathbf{i}},\tilde{\mathbf{j}}}_{\mu}} \langle \hat{S}^z_{\tilde{\mathbf{i}}}  \hat{S}^z_{\tilde{\mathbf{j}}}  \rangle$, $\mathcal{M}_2 = \sum^{\prime}_{\llangle \tilde{\mathbf{i}},\tilde{\mathbf{j}}\rrangle_{\text{diag}}} \langle \hat{S}^z_{\tilde{\mathbf{i}}}  \hat{S}^z_{\tilde{\mathbf{j}}} \rangle$ the summed correlations along nearest- and diagonal neighbors within the considered window in the bulk of squeezed space. Fig.~\ref{fig:recon_details}~\textbf{c} shows how approximations for $\mathcal{M}_1^{x,y}, \mathcal{M}_2$ quickly saturate with the number of used snapshots, suggesting a qualitatively satisfactory proxy for the spin-spin correlators after a few thousand projective measurements. For the rest of the analysis, we use sample sizes of 7,000 snapshots for each approximation of the correlations.

The minimization process is done via standard gradient descent (GD) methods, where in each iteration the parameters are updated according to the gradient $\nabla^{\prime} \mathcal{G}$ within the considered bulk window of squeezed space.
The temperature $\beta^{-1}$ of the $J_1^x-J_1^y-J_2$ Heisenberg Hamiltonian is chosen identically to the underlying simulations of the mixD $t-J$ system during the GD. Note that this choice might not reflect the actual effective temperature of the spin background. However, the relevant ratios $J_1^y/J_1^x$, $J_2/J_1^x$ that quantify the frustration in the system are independent of the true temperature of the squeezed magnetic environment. 

Intermediate temperature regimes $T/J \gtrsim 1$ -- as also chosen in our simulations -- have been shown to work best for reconstructions of the underlying coupling parameters, as both in the low and high temperature limit the energy landscape defined by $\mathcal{G}$ is entirely flat~\cite{anshu2021}.  Given a size $L_x\times L_y$ of the mixD system, the dimensions of the reconstructed $J_1^x-J_1^y-J_2$ Heisenberg ladder on the squeezed lattice is given by $\tilde{L}_x \times \tilde{L}_y = (1-n^h) L_x \times L_y$. 

Hamiltonian reconstruction results for a single run are presented in Fig.~\ref{fig:recon_details}~\textbf{d}. Evaluated correlations of the best fitting $J_1^x-J_1^y-J_2$ model are seen to perfectly match the measured mean correlations in the bulk of squeezed space, hence strongly supporting that the physics of the magnetic background in the mixD $t-J$ model is well captured by $J_1^x-J_1^y-J_2$ Heisenberg interactions on a square lattice. We have explicitly checked that independent of the initially chosen parameter values for the GD, $\{J_H\}$ always converge to identical points in parameter space, underlining the robustness of the GD scheme -- see Supplementary Note 1. 

To characterize and classify the reconstructed spin states in squeezed space as a function of doping, we perform ground state calculations of the $J_1^x-J_1^y-J_2$ Heisenberg model and evaluate the static spin structure factor (SSSF) given by $S(q_x, q_y) = \frac{1}{L_x^2 L_y^2 }\sum_{\mathbf{i},\mathbf{j}} e^{i \mathbf{q}\cdot(\mathbf{i} - \mathbf{j})}\braket{\hat{\mathbf{S}}_{\mathbf{i}} \cdot \hat{\mathbf{S}}_{\mathbf{j}}}$. For $J_1^x = J_1^y=J_1$, it has been demonstrated that a highly frustrated magnetic regime exists for $0.4 \lesssim J_2/J_1\lesssim 0.6$ that is sandwiched by a N\'eel and stripe AFM phase~\cite{Mezzacapo2012, Hu2013, Wang2013, Gong2014, Jiang2014, Haghshendas2018}. Though the exact nature of the non-magnetic ground state in the frustrated regime is still controversial, it remains a promising candidate for the realization of a quantum spin liquid phase possibly described by Anderson's resonating valence bond (RVB) paradigm~\cite{Anderson1973, Kivelson1987, Anderson1987, Kalmeyer1987, Kotliar1988, Nagaosa1990, Baskaran1993, White1994}. We evaluate the hybrid order parameter $S(\pi, \pi) + S(0, \pi)$ in the $J_1^y/J_1^x - J_2/J_1^x$ parameter space, signaling whether AFM or stripe AFM order exists in the system. Dark regions in the background of Fig.~\ref{fig:recon}~\textbf{b} correspond to no apparent spin ordering, and hence signal the existence of a strongly frustrated spin state akin to the observations in the homogeneous $J_1-J_2$ model.  

The reconstruction process, consisting of (i) approximating correlations in squeezed space using snapshots and (ii) performing the GD, is repeated a total number of ten times. Averaging over the converged results of all runs leads to the main result of this paper, presented in Fig.~\ref{fig:recon}~\textbf{b} by blue connected symbols. We observe how the spin state in squeezed space rapidly approaches the highly frustrated regime upon increasing the doping level, until seemingly saturating within it to a certain configuration $\{J^*_H\}$. We note that at the considered system sizes, boundary effects become especially pronounced at low hole concentrations $n^h = 0.05, 0.1$. This leads to a slow saturation of the correlations in the bulk of squeezed space, which in turn shifts effective couplings averaged within the fixed window to smaller (larger) values of $J_2/J_1^x$ ($J_1^y/J_1^x$), see Supplementary Note 2. In the thermodynamic limit, we in fact expect any finite hole doping in the chargon gas phase to drive the squeezed spin system into a highly frustrated state. A numerical study of longer ladders of size $40\times 2$ support this assumption, where already for $n^h = 0.05$ the spin state is reconstructed to lie deep inside the frustrated regime. 

The reconstruction scheme as introduced above takes into account spin-spin correlations directly measured in squeezed space, hence providing an unbiased platform for the analysis of the spin background by explicitly including the back-action of hole motion on the spins. Motivated from the separation of energy scales in the mixD $t-J$ model with $t/J \gg 1$, we make the ansatz of a fully decoupled thermal density matrix given by separate spin (s) and charge (c) sectors, $\hat{\rho} = \hat{\rho}_{\text{s}} \otimes \hat{\rho}_{\text{c}}$, and aim to test the resulting predictions against the unbiased reconstruction output.

Within the separation ansatz, interaction strengths in squeezed space are obtained by conditioned probabilities in real space~\cite{Hilker2017}. Two nearest neighbors along $x$ in squeezed space interact only if the corresponding sites are nearest neighbors along $x$ in real space, leading to an effective coupling strength (assuming homogeneous hole density $\braket{n^h_{\mathbf{i}}} = n^h$) -- see i.p. the supplementary materials of Ref.~\cite{Hilker2017},
\begin{equation}
J_1^x/J \propto \braket{(1-\hat{n}^h_{\mathbf{i}})(1-\hat{n}^h_{\mathbf{i} + \mathbf{e}_{x}})} = 1-2n^h + g^{(2)}_x.
\label{eq:pred_J1x}
\end{equation}
Here, $g^{(2)}_{\mu} = \braket{\hat{n}^h_{\mathbf{i}} \hat{n}^h_{\mathbf{i}+\mathbf{e}_{\mu}}}$ with $\mathbf{e}_{\mu}$ the unit vector in direction $\mu = x,y$.
Vertical and diagonal bonds are obtained similarly by conditioning the correlators by the total number of holes to the left of site $\mathbf{i}$, $\nu^h_{\mathbf{i} = [x,y]} = \sum_{x'<x} N^h_{[x',y]}$, with $N^h_{\mathbf{i}}$ the number of holes on site $\mathbf{i}$. Diagonal coupling strengths $J_n$ spanning a distance of $\Delta x = n-1$ (vertical bonds $J_1^y$ correspond to $J_1$ in this notation) are then given by
\begin{equation}
J_n/J \propto \braket{(1-\hat{n}^h_{\mathbf{i}})(1-\hat{n}^h_{\mathbf{i}+\mathbf{e}_{y}})}_{|\nu^h_{\mathbf{i}} - \nu^h_{\mathbf{i}+\mathbf{e}_y}|= n-1}.
\label{eq:pred_Jn}
\end{equation}

We evaluate the estimated effective couplings, Eqs.~\eqref{eq:pred_J1x} and~\eqref{eq:pred_Jn}, using the mixD $t-J$ snapshots. Results are shown by red connected symbols in Fig.~\ref{fig:recon}~\textbf{b}. We observe that the theoretically predicted expectations for $J_1^x, J_1^y, J_2$ within the separation ansatz agree remarkably well with the full reconstruction. Deviations from the above description, in particular the consistent underestimation of relative diagonal coupling strengths $J_2/J_1^x$, are likely caused by non-trivial spin-hole correlations in the mixD $t-J$ model, which are implicitly included in the reconstruction analysis but discarded in the separation ansatz. 

We note that the conditioned correlators Eqs.~\eqref{eq:pred_J1x} and~\eqref{eq:pred_Jn} calculated from mixD $t-J$ snapshots are numerically almost identical to calculations of free spinless fermions (free chargon gas), see Supplementary Note 3. This lets us conclude that holes, while behaving like free fermions in the chargon gas phase of the mixD $t-J$ model, nevertheless correlate non-trivially with the spin background.
\begin{figure}
\centering
\includegraphics[width=0.92\columnwidth]{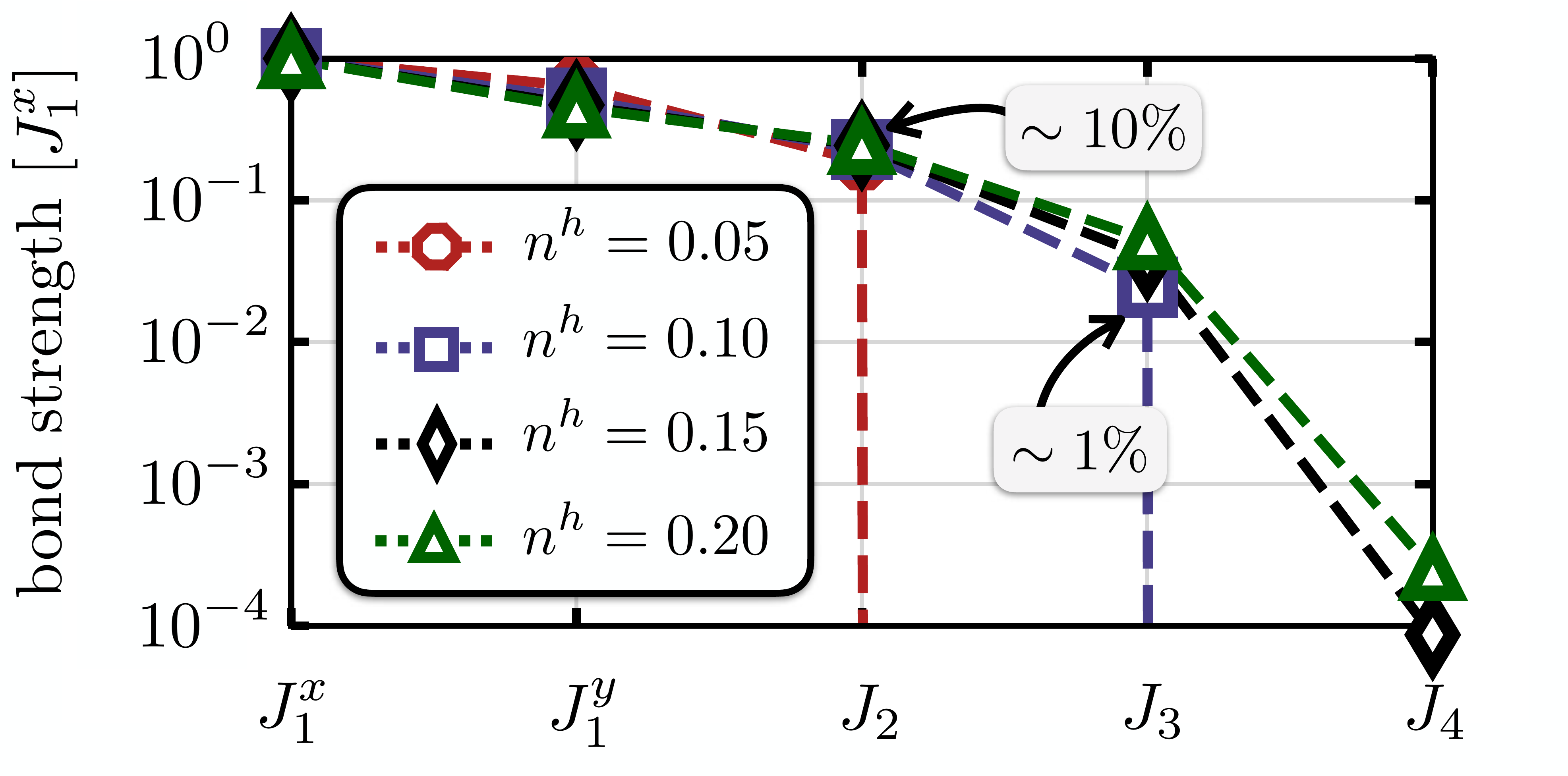}
\caption{\textbf{Significance of longer ranged couplings.} By evaluating the conditioned correlators, Eqs.~\eqref{eq:pred_J1x} and~\eqref{eq:pred_Jn}, we estimate the strengths of longer ranged couplings up to $J_4$. In units of the strongest interaction $J_1^x$, first order diagonal bonds $J_2$ are of the order of $\sim 10\%$, whereas couplings $J_3$ reach relative magnitudes of a few percent. Due to the finite system size, $J_3$ ($J_4$) and higher order couplings drop to zero for $n^h = 0.05$ ($n^h = 0.1$), as the corresponding conditioned probabilities Eq.~\eqref{eq:pred_Jn} vanish for a single (two) hole(s) per leg.}
\label{fig:LRC}
\end{figure}

So far, our approach has been to restrict the effective Hamiltonian Eq.~\eqref{eq:J1J2} to first order diagonal couplings $J_2$. To assess the systematic error related to this approximation, we estimate the magnitude of longer-range couplings $J_n$ by evaluating the conditioned correlators Eq.~\eqref{eq:pred_Jn} for $n\geq 3$. Relative strengths of couplings up to $J_4$ are depicted in Fig.~\ref{fig:LRC}, where a rapid decrease with real space distance is observed. The relative strength of $J_2$ in units of $J_1^x$ is of order $\sim 10\%$, cf. Fig~\ref{fig:recon}~\textbf{b}. $J_3$, on the other hand, reaches only a few percent in terms of $J_1^x$, suggesting that $n\geq 3$ couplings are negligible for the effective description. We implement the Hamiltonian reconstruction scheme outlined above also for a $J_1^x-J_1^y-J_2-J_3$ Heisenberg model including interactions in squeezed space up to $J_3$. We find that this results in only very minor corrections to the reconstructed ratios $J_1^y/J_1^x$, $J_2/J_1^x$, supporting that the spin physics is well captured by nearest-neighbor and frustrating (first order) diagonal bonds -- see Supplementary Note 4. \\

\begin{figure}
\centering
\includegraphics[width=0.93\columnwidth]{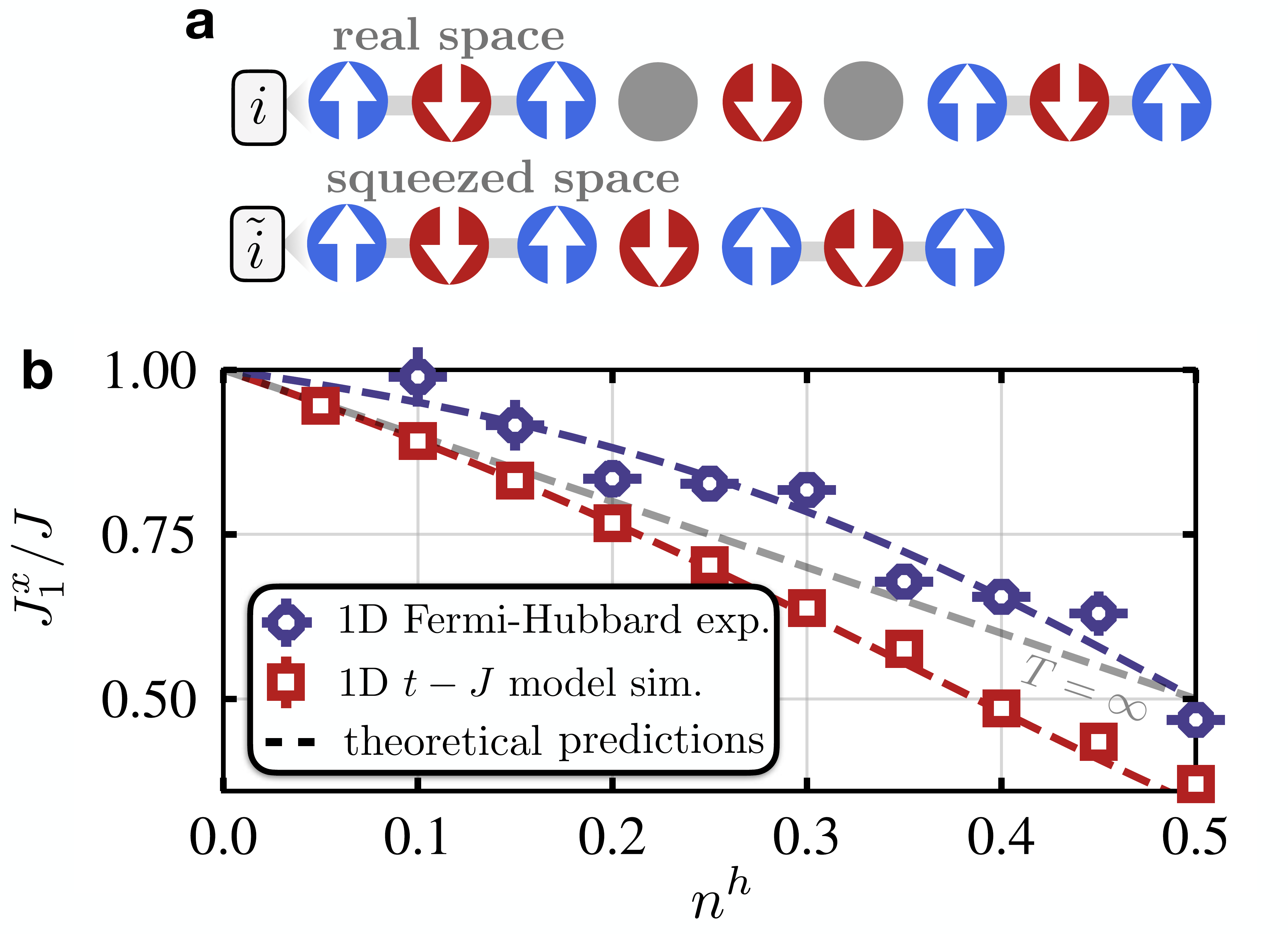}
\caption{\textbf{Reconstructing 1D systems from experiments.} \textbf{a}, Illustration of snapshots of the 1D FH model in real (top) and squeezed (bottom) space. \textbf{b}, Evaluation of 1D FH snapshots of a cold atom experiment~\cite{Hilker2017}. Reconstructions of the effective spin-Hamiltonian Eq.~\eqref{eq:Heis} in squeezed space for varying hole densities are shown by blue data points. Red data points correspond to reconstructions of the 1D $t-J$ model, which we simulate using MPS for parameters as estimated in~\cite{Hilker2017}, i.e., $t/J = 1.82$ and $T/J = 0.87$. Results are compared to theoretical predictions (dashed lines) assuming spin-charge separation, Eqs.~\eqref{eq:Jeff_tJ},~\eqref{eq:Jeff_FH}, showing a good match with the reconstructed data. In particular, higher order hopping processes lead to the FH measurement reconstructions of $J_1^x/J$ to consistently lie above predictions for the $t-J$ model. Error bars are too small to be visible for the $t-J$ reconstructions on the scale of the plot. The $T=\infty$ limit is shown by the grey dashed line, where a linear decrease $J_1^x = 1-n^h$ is expected for both the FH and $t-J$ model.}
\label{fig:1D}
\end{figure}

\textbf{Spin-charge separation in 1D.} In the 1D FH model, the ground state wave function is known to factorize into fully separated spin and charge channels in the strongly interacting limit, leading to the celebrated phenomenon of spin-charge separation (i.e., the exact absence of spin-hole correlations)~\cite{OgataShiba}. In 1D, is has been demonstrated that hidden spin correlations -- distorted in real space by the motion of holes -- can be revealed by transformation to squeezed space, effectively described by a 1D Heisenberg Hamiltonian with nearest neighbor interaction $J_1^x(n^h)$ on the squeezed lattice~\cite{Hilker2017, Coll_FH}, cf. Fig.~\ref{fig:1D}~\textbf{a},
\begin{equation}
\hat{\mathcal{H}}_{J_1^x} = J_1^{x} \sum_{\braket{\tilde{i},\tilde{j}}} \hat{\mathbf{S}}_{\tilde{i}} \cdot  \hat{\mathbf{S}}_{\tilde{j}}.
\label{eq:Heis}
\end{equation}

We apply our squeezed space Hamiltonian reconstruction scheme and recover the effective spin-Hamiltonian of the doped 1D FH model using experimentally obtained snapshots in a degenerate two-component ultracold Fermi gas carried out by some of us~\cite{Hilker2017}. Results are shown in Fig.~\ref{fig:1D}~\textbf{b} by blue data points, where a consistent decrease of effective coupling strength $J_1^x(n^h)$ is observed upon increasing the hole doping -- as expected from Eq.~\eqref{eq:pred_J1x}. For comparison, we further simulate the 1D $t-J$ model with identical parameters as estimated in the experiment ($t/J = 1.82$, $T/J = 0.87$) and use sampled thermal snapshots in squeezed space for reconstructions, shown by red squares in Fig.~\ref{fig:1D}~\textbf{b}. 

Effective interactions $J_1^x/J$ reconstructed for an underlying 1D $t-J$ model are seen to consistently lie below recovered coupling strengths of the 1D FH model. This discrepancy can be explained by higher order virtual processes in the FH model, which we illustrate by comparing the reconstructions to theoretical predictions within a separation ansatz. In the 1D $t-J$ model, effective spin interactions in squeezed space can be calculated via Eq.~\eqref{eq:pred_J1x}, yielding 
\begin{equation}
J_1^x/J = 1- n^h - \frac{1}{1-n^h} \left[G(1)\right]^2.
\label{eq:Jeff_tJ}
\end{equation}
Here, $G(d) = \frac{1}{\pi} \int_0^{\pi} dk \cos(kd) n_F(n^h, T)$ with $n_F(n^h, T)$ the Fermi-Dirac distribution of free chargons hopping on a 1D lattice at temperature $T$.
When generalizing the $t-J$ model to include next-nearest neighbor hole hopping processes mediated by doubly occupied virtual states as possible in the FH model, the effective coupling reads~\cite{Hilker2017}
\begin{equation}
    J_1^x/J = 1 - n^h + G(2),
    \label{eq:Jeff_FH}
\end{equation}
with $J = 4t^2/U$ and $U$ the Hubbard interaction. 

Reconstructed values of $J_1^x/J$ are observed to match the theoretical predictions for spin-charge separated systems well, depicted by red and blue dashed lines corresponding to Eqs.~\eqref{eq:Jeff_tJ} and~\eqref{eq:Jeff_FH}, respectively. This illustrates the (approximate) presence of spin-charge separation in the 1D FH and $t-J$ model away from the $T=0$ and strongly interacting limit, ultimately being mediated by their separation of energy scales. \\

\textbf{Discussion} 

Using Hamiltonian reconstruction schemes, we have proposed a method to quantify hole-motion-induced frustration in a doped antiferromagnet by exploiting the full information stored in many-body snapshots. 
An advantage of the reconstruction process as introduced above is that the effective Hamiltonian -- defined on the squeezed lattice -- describes a reduced number of degrees of freedom, i.e., its local Hilbert space dimension is smaller than the one of the original system. In particular, the effective spin-Hamiltonian in squeezed space is of local dimension $d=2$, rendering reconstructions for a given set of snapshots feasible even for larger system sizes. Experimental data of 2D systems that are inaccessible with classical simulations but within reach of current experiments~\cite{Hirthe2022} could be used as input for a computational reconstruction of the spin background. We benchmarked this for wide Heisenberg ladders, where we reconstruct unknown coupling parameters to high accuracy from snapshots, see Supplementary Note 5.  

By analyzing a setting in mixed-dimensions, we have firmly established a quantitative connection between the doped mixD $t-J$ model and the paradigmatic frustrated $J_1 - J_2$ model. In particular, we demonstrated how hole motion drives the spin background into a highly frustrated state, whereby effective diagonal, frustrating magnetic bonds are induced on the squeezed lattice formed by the spins alone. Our results match theoretical predictions based on spin-charge separation reasonably well, differences being likely caused by weak remaining spin-hole correlations that deserve further investigation in the future. We note that due to the formation of stripes below temperatures $T\sim J/2$~\cite{Schloemer2022} in the mixD $t-J$ model, its ground state is not directly related to a quantum spin liquid phase. Nevertheless, the ordered stripe phase at lower temperature may be merely covering a disordered quantum phase that dominates the physics of the model once the stripe order is melted away above the stripe critical temperature. Though this perspective is admittedly speculative, a similar view is often evoked in the context of describing the pseudogap with its associated small Fermi surface as originating from a covered ground state fractionalized Fermi liquid, see e.g. Ref.~\cite{Sachdev2020}.

Utilizing snapshots of a cold atom experiment simulating the 1D FH model, we already demonstrated the direct applicability of the reconstruction method to existing experimental data. From a converse experimental point of view, the above insights could further be utilized to effectively simulate the highly frustrated $J_1-J_2$ model by implementing the mixD setting and post-processing the measurements. 

The reconstruction scheme can be generalized and applied to a variety of many-body phases. In the stripe phase, for instance, fluctuating holes bound into stripes are expected to lead to spatial modulations of the couplings between spins which can be reconstructed using the scheme we described. Moreover, our method can be extended from mixD to fully 2D settings with homogeneous charge motion, where e.g. a weak easy-axis anisotropy of the Heisenberg interactions can enable string retracing~\cite{GrusdtX} to remove dominant charge fluctuations and define a squeezed lattice. Applying our scheme to such snapshots will provide a microscopic perspective on the doped FH model and its relation to putative topological order in his enigmatic model. Making explicit use of all accessible correlation functions in squeezed space to further enhance the accuracy of the reconstructions is a promising direction for future research, for instance by directly comparing the distributions of measured and reconstructed snapshots.  \\

\textbf{Methods}

\textbf{Finite-temperature DMRG.} We simulate the mixD $t-J$ model at finite temperature using mixed state purification schemes while conserving the system's symmetries~\cite{Nocera2016}. In particular, we expand the ladder system by introducing auxiliary sites, which act as a finite temperature bath via their entanglement to the physical system. In order to calculate thermal matrix product states, we first generate the infinite temperature, maximally entangled state $\ket{\Psi(\beta=0)}$ in a given symmetry sector~\cite{Schloemer2022} -- see also Supplementary Note 6.

A pure state in the enlarged system at finite temperature is then calculated by evolving $\ket{\Psi(\beta=0)}$ in imaginary time under the physical Hamiltonian, $\ket{\Psi(\tau)} = e^{-\tau \hat{\mathcal{H}}} \ket{\Psi(\beta=0)}$, where $\tau = \beta/2$ with $\beta$ the inverse temperature. The corresponding mixed state of the physical system is computed by tracing out all auxiliary degrees of freedom when computing expectation values in the physical subset.

During the imaginary time evolution, we conserve the particle number in each physical leg $N_{\ell}, \ell=1..L_y$, the total particle number in the auxiliary system $N_{\text{aux.}}^{\text{tot}}$, as well as the total spin $S^{z,\text{tot}}_{\text{phys.+aux.}}$ (the latter allowing for finite total magnetizations of the physical system at finite temperate). This results in a total of $L_y + 2$ symmetries employed by the DMRG implementation.

Given a generic observable $\hat{O}$ of the physical chain, the thermodynamic average can be calculated in the enlarged space by tracing out the ancilla degrees of freedom, \begin{equation}
    \braket{\hat{O}}_{\beta} = \frac{\braket{\Psi(\beta)|\hat{O}|\Psi(\beta))}}{\braket{\Psi(\beta)|\Psi(\beta)}}.
\end{equation}
Here, the norm $\braket{\Psi(\beta)|\Psi(\beta)} \propto Z(\beta)$ is proportional to the partition function at temperature $\beta^{-1}$. 

The maximally entangled state needed as a starting point of the imaginary time evolution is generated using the concept of entangler Hamiltonians~\cite{Feiguin2010, Nocera2016}, which we specifically tailor for our ``leg-canonical'' ensemble~\cite{Schloemer2022}. Since the maximally entangled state is usually of low bond dimension, we first employ global MPS imaginary time evolution schemes to evolve the system away from infinite temperature. Once bond dimensions are sufficiently high, we switch to local approximation methods. In particular, we use the Krylov scheme and the time-dependent variational principle (TDVP) for global and local evolutions, respectively~\cite{Paeckel_time}.

\textbf{Hamiltonian reconstruction.} Due to the non-Markovian nature of quantum states, it is \textit{a priori} unclear whether measured correlations are sufficient to learn the quantum interactions of the underlying Hamiltonian~\cite{Leifer2008}. However, it has been shown that the strongly convex property of the free energy with respect to the interaction parameters renders the Hamiltonian learning problem feasible~\cite{anshu2021}. 

In each GD step, we compute the partition function and relevant correlations $\mathcal{M}_1^{x/y}, \mathcal{M}_2$ using the MPS schemes described above. Due to the numerical complexity, we do not consider advanced GD methods with varying step size e.g. given by the Amijo rule, but stick to a straightforward optimization using a fixed descent step. In particular, we choose the step $\mathbf{a}$ to be 20 \% of the objective gradient, i.e. $\textbf{a} = 0.2 \nabla^{\prime} \mathcal{G}$, where $\nabla^{\prime}$ is the gradient in parameter space within the fixed window in the bulk of squeezed space as introduced in the main text. When the norm of the gradient reaches a certain threshold, here chosen as $|\nabla^{\prime} \mathcal{G}| < 10^{-6}$, we stop the descent and assume converged results. 

In our simulations, we work at intermediate temperatures, i.p. $T/J=5/3$. On the one hand, this ensures that the mixD $t-J$ system is in the chargon gas phase, i.e., stripes do not form~\cite{Schloemer2022}. On the other hand, intermediate temperature regimes have been shown to yield best reconstruction results from projective measurements~\cite{anshu2021}. To illustrate the latter argument, consider for instance the ferromagnetic (FM) Ising model, featuring a FM ground state for any non-zero interaction strength. Therefore, at low temperatures close to the ground state, the energy landscape $\mathcal{G}$ is nearly flat, resulting in bad reconstructions. Similarly, for $T/J \gg 1$ measured correlations only weakly depend on the underlying coupling parameters (e.g. the infinite temperature state is identical for all interaction strengths), which hinder precise reconstructions. By reconstructing purely magnetic models for various temperatures, we demonstrate this explicitly in Supplementary Note 5. \\

\textbf{Data availability} 

The datasets generated and/or analysed during the current study are available from the corresponding author on reasonable request. \\

\textbf{Code availability} 

The data analysed in the current study has been obtained using the SyTen package~\cite{hubig:_syten_toolk, hubig17:_symmet_protec_tensor_networ}. \\

\textbf{Author Contributions} 

A.B., F.G. and H.S. conceptualized the idea. H.S. carried out the calculations under the supervision of A.B., F.G. and U.S. T.A.H. and I.B. supervised the application to the experimental data. All authors discussed the results. H.S. wrote the manuscript with input from all authors. \\

\textbf{Competing interests}

The authors declare no competing interests. \\

\textbf{Acknowledgments} 

We are thankful for valuable discussions with F. Palm, M. Kebri\v{c}, and L. Homeier. This research was funded by the Deutsche Forschungsgemeinschaft (DFG, German Research Foundation) under Germany’s Excellence Strategy—EXC-2111—390814868, by the European Research Council (ERC) under the European Union’s Horizon 2020 research and innovation programme (grant agreement number 948141) – ERC Starting Grant SimUc-Quam, and by the NSF through a grant for the Institute for Theoretical Atomic, Molecular, and Optical Physics at Harvard University and the Smithsonian Astrophysical Observatory. \\

\appendix
\widetext

\section{\underline{Supplementary Materials}}

\setcounter{equation}{0}
\setcounter{figure}{0}
\setcounter{table}{0}
\makeatletter
\renewcommand{\theequation}{S\arabic{equation}}
\renewcommand{\thefigure}{S\arabic{figure}}

\subsection{Supplementary Note 1: Convergence of effective couplings}

\begin{figure}[h]
\centering
\includegraphics[width=0.8\columnwidth]{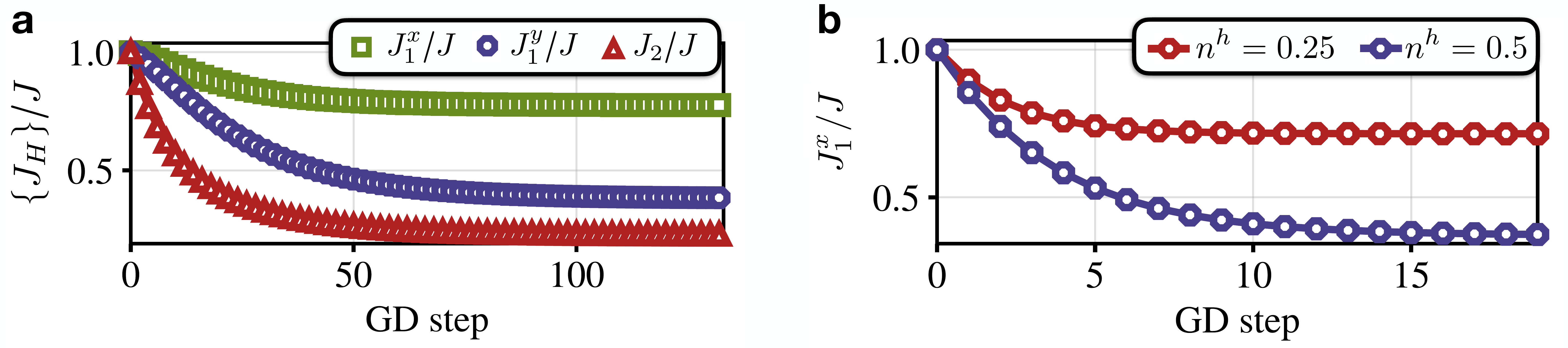}
\caption{\textbf{Convergence during GD.} \textbf{a}, A single GD path for the couplings $J_1^x/J, J_1^y/J, J_2/J$ in the mixD $t-J$ model at $n^h=0.2$ with parameters as chosen in Fig.~2. The GD is initialized with parameters $J_1^x/J = J_1^y/J = J_2/J = 1$. Only every second data point is shown for illustrative purposes. \textbf{b}, Convergence of the effective coupling $J_1^x/J$ in the 1D $t-J$ model for $n^h = 0.25, 0.5$, with parameters as chosen in Fig.~4. We initialize $J_1^x/J = 1$.}
\label{fig:path_J}
\end{figure}

During the gradient descent scheme introduced in the main text, effective parameters describing the spin state in squeezed space are updated in each iteration until correlations of the reconstructed model match the measured correlations in squeezed space. Fig.~\ref{fig:path_J}~\textbf{a} illustrates paths of the effective couplings $J_1^x/J, J_1^y/J, J_2/J$ during the GD process in case of the doped mixD $t-J$ model, cf. Fig.~2~\textbf{d} in the main text. The step $\mathbf{a}$ is chosen to to be 20 \% of the objective gradient, i.e. $\textbf{a} = 0.2 \nabla^{\prime} \mathcal{G}$, cf. Eq.~(3) in the main text.

After initialization at $J_1^x/J = J_1^y/J = J_2/J = 1$, the parameters are observed to converge after $\sim 100$ GD steps. For the 1D $t-J$ model (cf. Fig.~4), we observe similar convergence after $\sim 15$ updates of the effective parameter, as exemplified by a single GD path of the coupling $J_1^x/J$ in Fig.~\ref{fig:path_J}~\textbf{b}.
We have explicitly verified that the converged results are independent of the initially chosen parameters, underlining the robustness of the GD scheme when reconstructing effective Hamiltonians from snapshots. 

\subsection{Supplementary Note 2: Finite size effects}

We have argued in the main text that the observed flow into the highly frustrated regime when increasing the doping level  (cf. Fig.~1~\textbf{b}) results from the finite size of the simulated ladders, and that in the thermodynamic limit we in fact expect any hole doping $n^h >0$ to drive the spin background into a frustrated state. Fig.~\ref{fig:reconJ3} shows correlations in squeezed space for all considered hole dopings $n^h = 0.05, 0.1, 0.15, 0.2$ in analogy to~Fig.~2~\textbf{b}.

\begin{figure}[h]
\centering
\includegraphics[width=0.9\columnwidth]{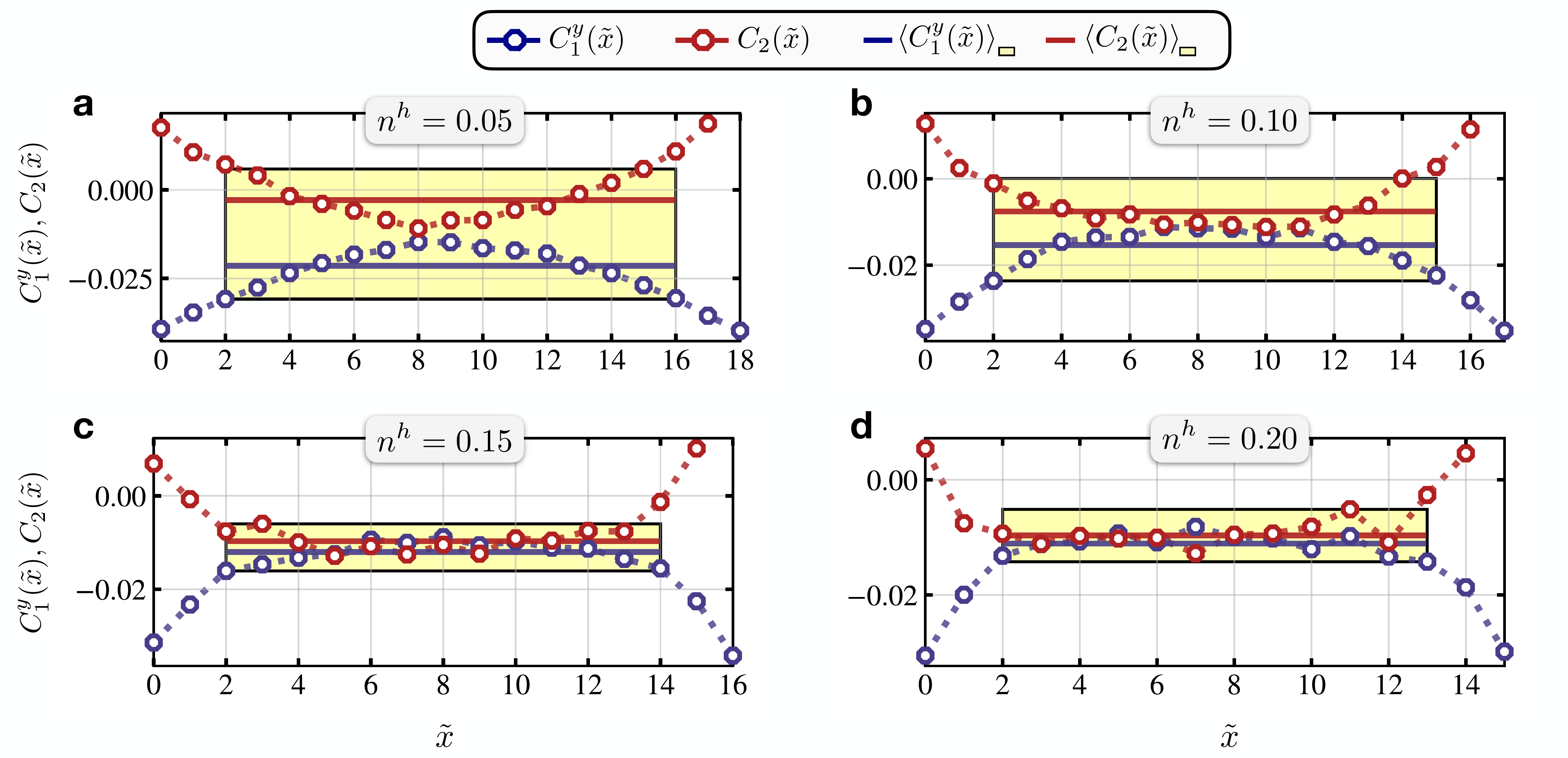}
\caption{\textbf{Spin-spin correlations in squeezed space.} Correlations $C_1^y(\tilde{x}) = \braket{\hat{S}^z_{[\tilde{x},\tilde{0}]} \hat{S}^z_{[\tilde{x},\tilde{1}]}}$ and $C_2(\tilde{x}) =  \braket{\hat{S}^z_{[\tilde{x},\tilde{0}]} \hat{S}^z_{[\tilde{x}+1,\tilde{1}]}} + \braket{\hat{S}^z_{[\tilde{x}+1,\tilde{0}]} \hat{S}^z_{[\tilde{x},\tilde{1}]}}$ for $n^h = 0.05, 0.1, 0.15,0.2$. Parameters are chosen identically to Fig.~1 in the main text. Solid lines show averaged values over the window when discarding the outermost two sites. Correlations $C_1^y(\tilde{x})$ ($C_2(\tilde{x})$) are seen to strengthen (weaken) by averaging over non-saturated configurations.}
\label{fig:reconJ3}
\end{figure}

Solid lines illustrate averaged correlations in the fixed bulk window in squeezed space by discarding the outer two sites. For low hole concentrations, $n^h = 0.05,0.1$, boundary effects are particularly noticeable. Correlations in squeezed space only slowly saturate, which which in turn shifts effective couplings averaged within the fixed bulk window in squeezed space to smaller (larger) values of $J_2/J_1^x$ ($J_1^y/J_1^x$). This effect ultimately results in the observed flow into the highly frustrated region when increasing hole doping as discussed in the main text. 

\begin{figure}[h]
\centering
\includegraphics[width=0.9\columnwidth]{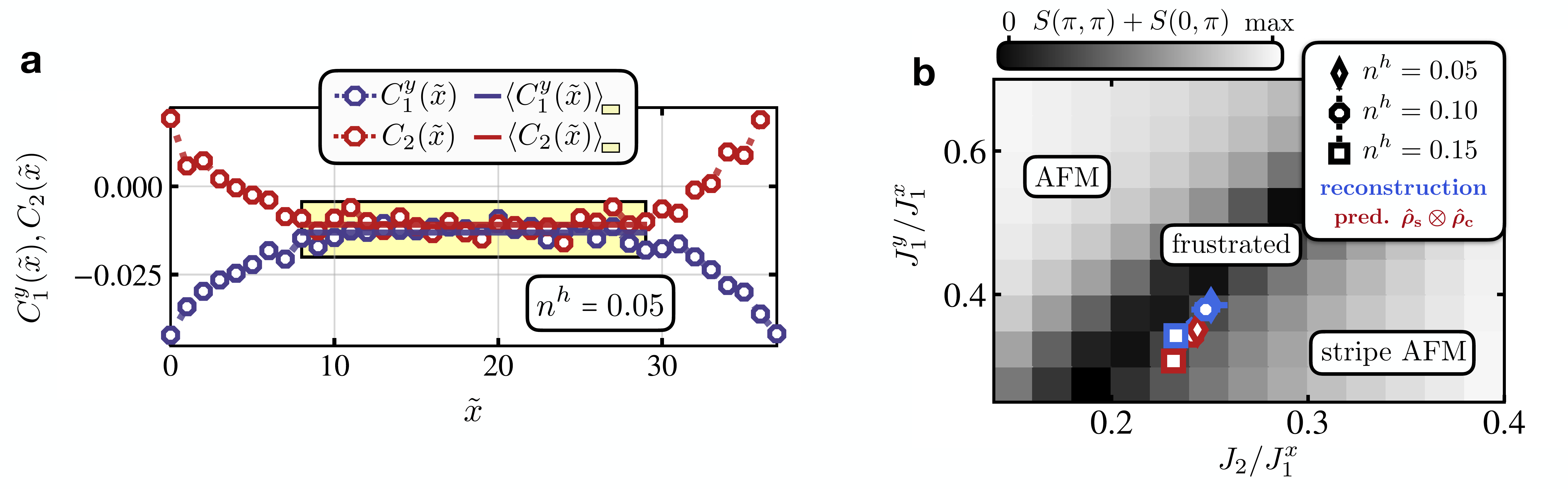}
\caption{\textbf{Reconstruction of long ladders.} \textbf{a}, Correlations $C_1^y(\tilde{x}) = \braket{\hat{S}^z_{[\tilde{x},\tilde{0}]} \hat{S}^z_{[\tilde{x},\tilde{1}]}}$ and $C_2(\tilde{x}) =  \braket{\hat{S}^z_{[\tilde{x},\tilde{0}]} \hat{S}^z_{[\tilde{x}+1,\tilde{1}]}} + \braket{\hat{S}^z_{[\tilde{x}+1,\tilde{0}]} \hat{S}^z_{[\tilde{x},\tilde{1}]}}$ in squeezed space for a $40 \times 2$ mixD $t-J$ ladder, with $t/J = 3$, $n^h = 0.05$ at $T/J=5/3$ and using 20,000 snapshots. In the bulk of squeezed space, saturation of the correlators is observed for all considered hole dopings, as illustrated by the solid lines corresponding to averages within the boxed region. \textbf{b}, Hamiltonian reconstruction results reveal that for all hole concentrations, hole motion drives the spins into a highly frustrated state (blue symbols). We discard the outer eight sites in squeezed space when reconstructing the effective Hamiltonians, as illustrated by the yellow box in \textbf{a}. The separation ansatz as discussed in the main text yields similar results as the full reconstruction, illustrated by red symbols.}
\label{fig:40by2}
\end{figure}

To underline the former arguments, we simulate a long ladder of size $40 \times 2$, with $t/J=3$ and at $T/J = 5/3$. Correlations in squeezed space are shown in Fig.~\ref{fig:40by2}~\textbf{a} for $n^h = 0.05$. When averaging the correlations in the bulk of squeezed space, we discard the outer eight sites, as illustrated by the yellow box. The larger system size leads to saturated correlations in the considered bulk window also for low hole concentrations.

After reconstructing effective $J_1^x-J_1^y-J_2$ Hamiltonians for the long ladder, we observe that even for small hole doping, $n^h=0.05$, the spin system is reconstructed to lie inside the frustrated regime, as illustrated by blue symbols in Fig.~\ref{fig:40by2}~\textbf{b}. We again observe that a separation ansatz $\hat{\rho} = \hat{\rho}_{\text{c}} \otimes \hat{\rho}_{\text{s}}$ yields similar qualitative results as the full reconstruction, cf. the red symbols in Fig.~\ref{fig:40by2}~\textbf{b}.

\subsection{Supplementary Note 3: Free chargon gas}
We have seen in the main text that effective coupling parameters calculated within a separation ansatz, $\hat{\rho} = \hat{\rho}_{\text{c}} \otimes \hat{\rho}_{\text{s}}$, agree remarkably well with the full reconstruction results. The conditioned correlators, Eq.~(4) and (5), were calculated using simulated snapshots of mixD ladders. When taking snapshots of free fermions hopping on a lattice (free chargon gas), evaluation of the same hole-hole correlators yield numerically almost indistinguishable results from the mixD snapshots. 
\begin{figure}[h]
\centering
\includegraphics[width=0.4\columnwidth]{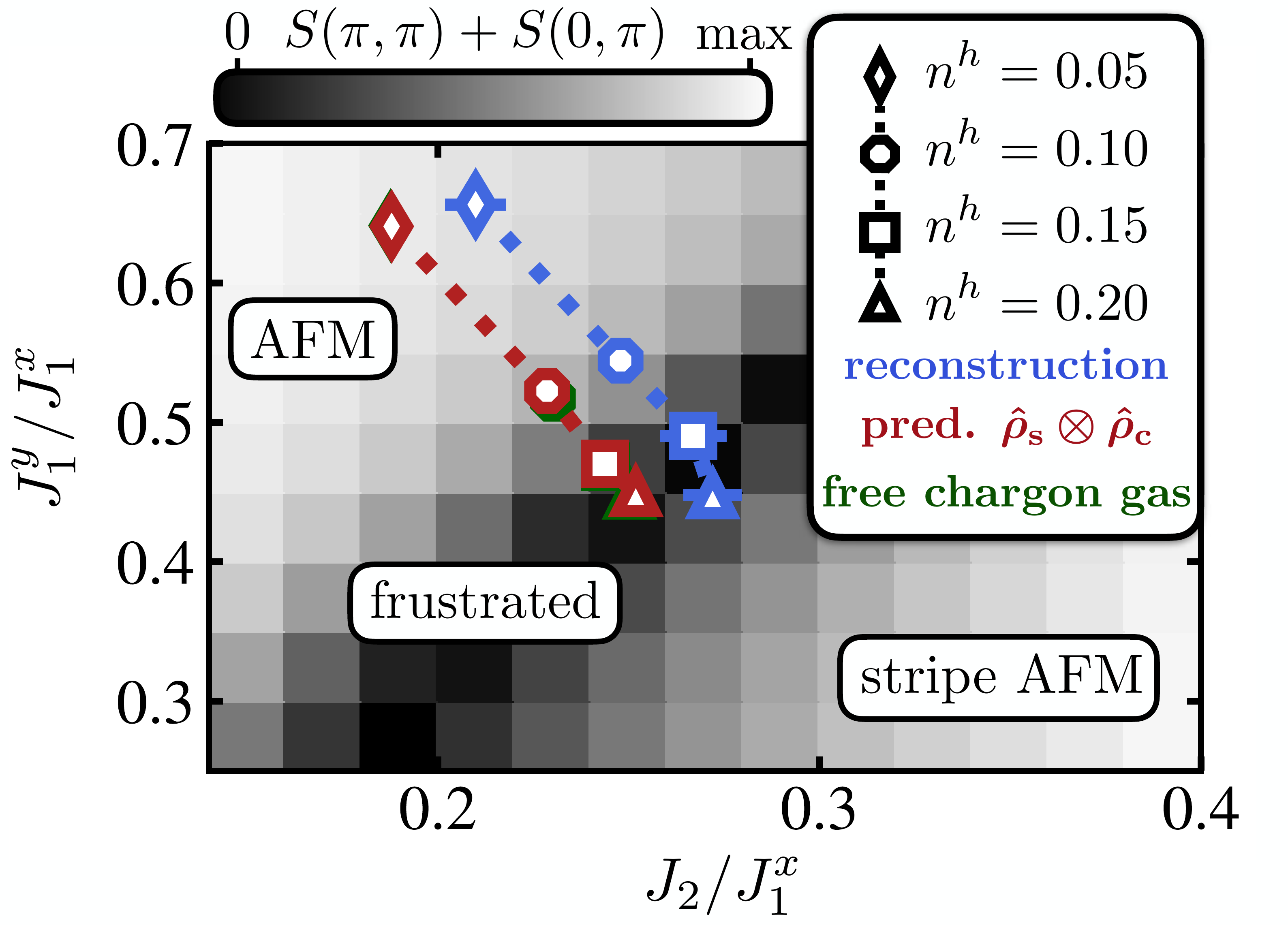}
\caption{\textbf{Comparison of Hamiltonian reconstruction with theoretical predictions.} Hamiltonian reconstruction results (blue) for hole dopings $n^h = 0.05 \dots 0.2$ of a mixD $t-J$ ladder with $t/J = 3$, $L_x \times L_y = 20 \times 2$ and $T/J = 5/3 \sim 1.67$. Reconstructions of squeezed space according to the input $J_1^x-J_1^y-J_2$ Heisenberg Hamiltonian, Eq.~(2), are presented in a $J_1^y/J_1^x, J_2/J_1^x$ diagram. Red connected symbols show theoretical expectations assuming no spin-hole correlations in the mixD $t-J$ model, i.e., $\hat{\rho} = \hat{\rho}_{\text{s}} \otimes \hat{\rho}_{\text{c}}$, evaluated via Eq.~(4) and~(5). Green data points show evaluation of Eq.~(4) and~(5) for a free chargon gas, being almost indistinguishable from the mixD calculations. Blue and red curves are identical to the data presented in Fig.~1.}
\label{fig:comp_free}
\end{figure}
This is illustrated in Fig.~\ref{fig:comp_free}, where effective parameters $J_1^x, J_1^y, J_2$ are estimated both from mixD (red) as well as free chargon (green) snapshots. The two sets of data points are barely distinguishable from another, which lets us conclude that deviations from the full reconstructions are likely due to remaining non-trivial spin-hole correlations in the mixD system.

\subsection{Supplementary Note 4: Longer-range couplings}
In the main text, spin states of mixD $t-J$ ladders in squeezed space were reconstructed using an effective $J_1^x-J_1^y-J_2$ Heisenberg model, i.e., first order next-nearest neighbor couplings were taken into account via diagonal magnetic bonds. Though longer-range couplings $J_n$, $n\geq 3$ also enter the effective descriptions of spins in squeezed space (Fig.~\ref{fig:reconJ3}~\textbf{a}), the approximation of neglecting higher order couplings was justified in the main text by approximating their weight via Eq.~(4) and~(5).

\begin{figure}[h]
\centering
\includegraphics[width=0.85\columnwidth]{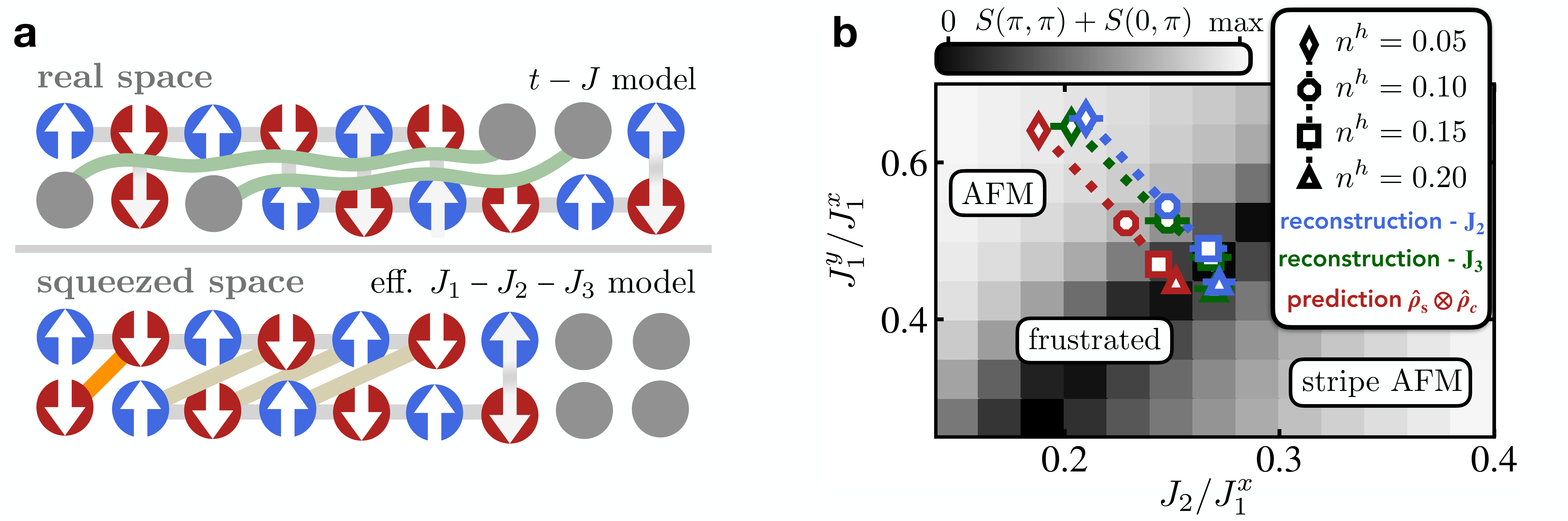}
\caption{\textbf{Hamiltonian reconstruction including longer-range couplings.} \textbf{a}, Schematic illustration of how longer-range bonds appear in the effective magnetic system in squeezed space. \textbf{b}, Reconstruction results when including terms up to $J_2$ (blue connected symbols) and $J_3$ (green connected symbols). Though small corrections are visible, qualitative results in comparison to the separation ansatz (red connected symbols) remain unchanged.}
\label{fig:reconJ3}
\end{figure}

To underline this, we reconstruct the effective spin-Hamiltonian from measurements in squeezed space, here by explicitly including bonds up to $J_3$ on the squeezed lattice. The corresponding $J_1-J_2-J_3$ type Hamiltonian reads
\begin{equation}
\hat{\mathcal{H}}_{J_1^x, J_1^y, J_2, J_3} = \sum_{\mu = x,y} J_1^{\mu} \sum_{\braket{\tilde{\mathbf{i}},\tilde{\mathbf{j}}}_{\mu}} \hat{\mathbf{S}}_{\tilde{\mathbf{i}}} \cdot  \hat{\mathbf{S}}_{\tilde{\mathbf{j}}} + J_2 \sum_{\llangle \tilde{\mathbf{i}},\tilde{\mathbf{j}} \rrangle_{\text{diag}_2}}  \hat{\mathbf{S}}_{\tilde{\mathbf{i}}} \cdot  \hat{\mathbf{S}}_{\tilde{\mathbf{j}}} + J_3 \sum_{\llangle \tilde{\mathbf{i}},\tilde{\mathbf{j}} \rrangle_{\text{diag}_3}}  \hat{\mathbf{S}}_{\tilde{\mathbf{i}}} \cdot  \hat{\mathbf{S}}_{\tilde{\mathbf{j}}}.
\label{eq:J1J2J3}
\end{equation}

Reconstruction results are presented in Fig.~\ref{fig:reconJ3}~\textbf{b}. We observe that corrections of the reconstructed parameter values when considering terms up to $J_3$ are very minor -- underlining how the physics in squeezed space is well captured by nearest-neighbor and diagonal frustrating couplings, i.e., $J_1^x, J_1^y, J_2$. 

\subsection{Supplementary Note 5: Reconstruction of multi-leg ladders.} 

In the main text, it was argued that the reduced number of degrees of freedom in pure spin-Hamiltonians that are needed for reconstructions of e.g. the FH or $t-J$ model make an analysis of experimentally measured 2D systems conceivable with state-of-the-art numerical techniques. To demonstrate the ability to reconstruct such wide, 2D-like geometries, we consider a $J_1^x-J_1^y-J_2$ Heisenberg model of size $L_x\times L_y = 20 \times 4$. From MPS simulations we take snapshots of the thermal state, which we utilize to estimate spin-spin correlations, and aim to reconstruct the original Hamiltonian from the measurements. We choose $J_1^y/J_1^x = 0.7$, $J_2/J_1^x=0.3$, and $T/J_1^x = 5/3$.
\begin{figure}[h]
\centering
\includegraphics[width=0.55\columnwidth]{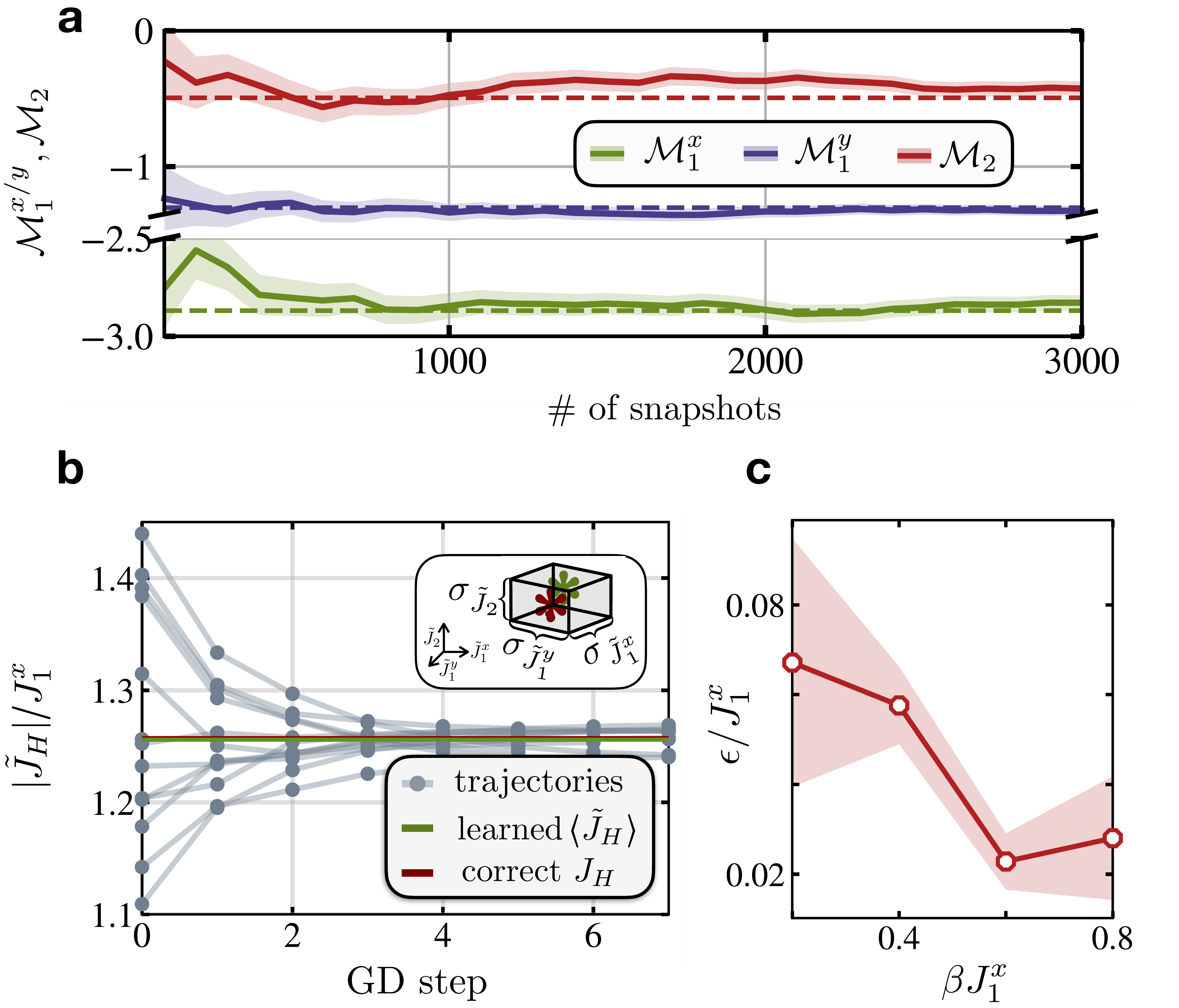}
\caption{\textbf{Reconstructing multi-leg systems.} \textbf{a}, Approximation of the correlations $\mathcal{M}_1^{x,y}, \mathcal{M}_2$ with randomly sampled snapshots (solid lines) together with an exact evaluation using the underlying thermalized MPS (dashed lines). Errors to the mean are indicated by the coloured background. The $J_1^x-J_1^y-J_2$ system is of size $L_x\times L_y = 20 \times 4$, with $J_1^y/J_1^x = 0.7$, $J_2/J_1^x = 0.3$, and $T/J_1^x = 5/3$. \textbf{b}, Learning the Hamiltonian's coupling parameters using GD. We start the GD from randomized positions in coupling space and plot the resulting paths of $|\tilde{J}_H| = \sqrt{(\tilde{J}_1^x)^2 + (\tilde{J}_1^y)^2 + (\tilde{J}_2)^2}$ during the GD. For each run, correlations are approximated using 3,000 measurements. The actual Hamiltonian's parameters are marked by the red solid line. The green line results from averaging $|\tilde{J}_H|$ after 7 GD steps. The inset shows in more detail the resulting averages of $\{\tilde{J}_H\}$ (green asterisk), with their corresponding errors (black box). For all three parameters, the actual parameters (red asterisk) match the predicted ones. \textbf{c}, $\ell_2$ error of the learned and actual parameters for a $12\times 2$ system after reaching convergence for varying temperature $\beta J_1^x \sim0.2 \dots 0.8$. Errors to the mean are indicated by the coloured background. Intermediate temperatures are found to work best for a reconstruction.}
\label{fig:recon_Heis}
\end{figure}

In Fig.~\ref{fig:recon_Heis}~\textbf{a}, we see how the measurement outcomes using snapshot sample sets of various size (solid lines) converge towards the exact MPS result of the simulated system (dashed lines) after a few thousand measurements. 

We again initialize the GD with randomly chosen parameters and stop the descent once convergence is reached, which we repeat multiple times to estimate statistical errors. Fig.~\ref{fig:recon_Heis}~\textbf{b} shows GD trajectories of the absolute value of the three parameters, $|\tilde{J}_H| = \sqrt{(\tilde{J}_1^x)^2 + (\tilde{J}_1^y)^2 + (\tilde{J}_2)^2}$, for seven repetitions. We observe how each trajectory rapidly flows into the region of the correct (absolute value) of interaction parameters, illustrated by the red solid line in Fig.~\ref{fig:recon_Heis}~\textbf{b}. By averaging the final parameters of all trajectories after seven GD steps, the estimated couplings are seen to all lie within one standard deviation to the correct interaction strengths, cf. the inset in Fig.~\ref{fig:recon_Heis}~\textbf{b}. We have verified that the descent converges to identical values independent of the initially chosen parameters -- including initial points lying very far away from the true values.

As already mentioned in~\cite{anshu2021}, the reconstruction accuracy depends on the system's temperature. We repeat the process for several different temperatures $\beta J_1^x=0.2,0.4,0.6,0.8$ and measure the $\ell_2$ norm between the true $\{J_H\}$ and learned $\{\tilde{J}_H\}$ interaction strengths, $\epsilon = [\sum_{\mu = x,y} (J_1^{\mu} - \tilde{J}_1^{\mu})^2 + (J_2 - \tilde{J_2})^2]^{1/2}$, again for multiple repetitions and after achieving GD convergence. The resulting $\ell_2$ error as a function of inverse temperature is shown in Fig.~\ref{fig:recon_Heis}~\textbf{c}. We observe that reconstructions work best around intermediate temperatures $\beta J_1^x \sim 0.6$, which \textit{a posteriori} justifies our choice of temperature for the simulation and reconstruction of the mixD ladders and, furthermore, is in a temperature regime accessible for quantum gas microscopes. \\

\subsection{Supplementary Note 6: Numerical finite temperature simulations}

We use purification schemes within the MPS framework to simulate both mixD $t-J$ and $J_1^x-J_1^y-J_2$ Heisenberg models at finite temperature. In the spin sector, we employ a  grand-canonical ensemble, allowing for thermal spin magnetizations. In the mixD model, we conserve its additional symmetries in the charge sector, namely the separate $U(1)$ charge conservation symmetries in each ladder leg (i.e. in the charge sector, calculations are canonical in each leg separately).

We here shortly summarize the purification schemes used in the main text, and refer to e.g.~\cite{Nocera2016, Schloemer2022} for more detailed discussions. When purifying the system, the Hilbert space is enhanced by one auxiliary (often also called ancilla) site per physical site, which allows to display mixed states in the physical subset of the Hilbert space as pure states on the enlarged space. Thermal expectation values are then computed via
\begin{equation}
    \braket{\hat{O}}_{\beta} = \frac{\braket{\Psi(\beta)|\hat{O}|\Psi(\beta)}}{\braket{\Psi(\beta)|\Psi(\beta)}},
\end{equation}
where $\ket{\Psi(\beta)} = e^{-\beta \hat{H}/2} \ket{\Psi(\beta=0)}$ is the maximally entangled state $\ket{\Psi(\beta=0)}$ evolved in imaginary time $\tau = \beta/2$, and $\mathcal{O}$ is an operator acting on the physical sites only (i.e., all auxiliary degrees of freedom are traced out in the evaluation of the expectation value). Note that this is indeed an exact formulation of the usual form $\braket{\hat{\mathcal{O}}} = \frac{1}{Z} \text{Tr} ( \rho \hat{O} )$, where $Z = \text{dim}(\mathcal{H}) \braket{\Psi(\beta)|\Psi(\beta)}$ with $\text{dim} (\mathcal{H})$ the dimension of the Hilbert space. 

\begin{figure}[h]
\centering
\includegraphics[width=0.5\columnwidth]{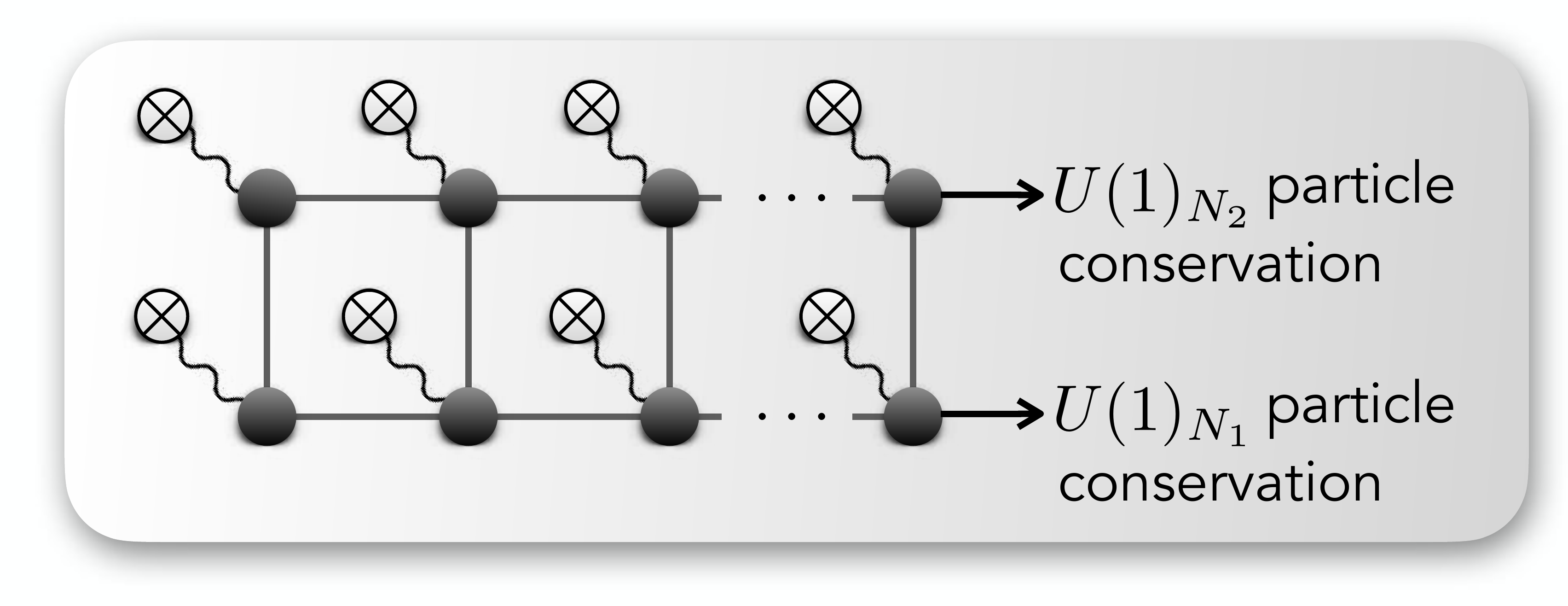}
\caption{Cartoon of the physical and ancilla system to emulate a thermal bath for finite temperature calculations. Physical sites and bonds are illustrated by grey filled circles and black lines, auxiliary sites and their artificial connection to the physical system are shown by crosses and wavy lines. There exists no physical connection between the two Hilbert spaces, nevertheless ancilla sites can act on physical sites implicitly through their entanglement.}
\label{fig:ancilla}
\end{figure}

The structure of the system employed in our calculations is depicted in Fig.~\ref{fig:ancilla}. Physical sites (interactions) are represented by solid circles (lines), ancilla sites are shown by crossed circles connected to their corresponding physical sites by wavy lines. 

The maximally entangled state for the Heisenberg model at infinite temperature and in the grand-canonical ensemble reads
\begin{equation}
\label{eq:max_ent_heis}
    \ket{\Psi(\beta=0)} = \bigotimes_{i = 0}^{L-1} \left( \sum_{\sigma=\uparrow,\downarrow} \ket{\sigma, \bar{\sigma}} \right).
\end{equation}
Here, $\{\ket{\uparrow}, \ket{\downarrow}\}$ is the single particle basis of the Heisenberg model with $\bar{\uparrow} = \downarrow$, $\bar{\downarrow} = \uparrow$.

\begin{figure}
\centering
\includegraphics[width=0.6\columnwidth]{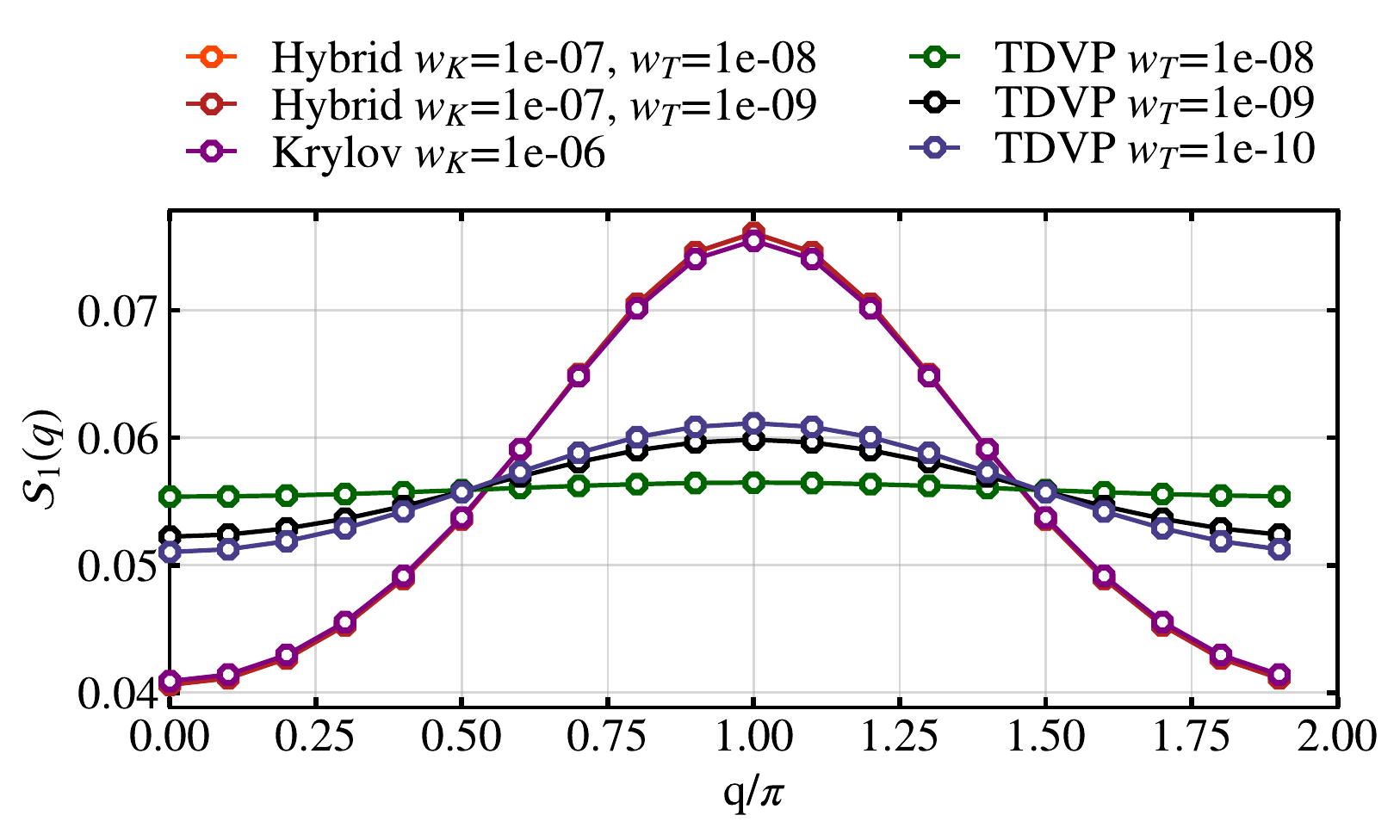}
\caption{Convergence properties of finite-T simulations of a $L_x \times L_y = 40 \times 4$ $J_1^x - J_1^y - J_2$ Heisenberg model. To evaluate convergence of spin-spin correlations, we compute the static spin-structure factor (SSSF) along a single ladder leg $y$, given by $\mathcal{S}_y(q) = \frac{1}{\sqrt{L_x}}\sum_{x_i,x_j} e^{-i q (x_i-x_j)} \braket{\hat{S}^z_{[x_i,y]} \hat{S}^z_{[x_j,y]}}$. Purely local TDVP evolution schemes starting from the maximally entangled state, Eq.~\eqref{eq:max_ent_heis}, perform poorly compared to global evolution methods (Krylov). A single global step with $\Delta \tau = 0.1$ followed by TDVP imaginary time evolution (denoted by ``Hybrid'') is observed to be sufficient to reach convergence towards to the fully global result.}
\label{fig:conv}
\end{figure}

On the other hand, in the mixD system, whose local Hilbert space dimension is larger by one degree, we have for the maximally entangled state in the ``leg-canonical'' ensemble in the charge sector and the grand-canonical ensemble in the spin sector
\begin{equation}
\label{eq:max_ent}
    \ket{\Psi(\beta=0)} = \prod_{\ell} \hat{\mathcal{P}}_{N_{\ell}} \bigotimes_{i = 0}^{L-1} \left( \ket{0,0} + \sum_{\sigma=\uparrow,\downarrow} \ket{\sigma, \bar{\sigma}} \right).
\end{equation}
Here, $\ell$ is the physical chain index, $L= L_x L_y$ the total number of physical sites in the ladder system, $\{\ket{0}, \ket{\uparrow}, \ket{\downarrow}\}$ is the single particle basis of the $t-J$ model with $\bar{\uparrow} = \downarrow$, $\bar{\downarrow} = \uparrow$, and $\hat{\mathcal{P}}_{N_{\ell}}$ is the projector to the subspace with $N_{\ell}$ dopants in the $\ell^{\text{th}}$ physical chain; the first and second entries in the kets correspond to physical and auxiliary sites, respectively. \\

In order to get the MPS representation of the maximally entangled state, we perform a ground state search of specifically tailored entangler Hamiltonians, cf.~\cite{Nocera2016} for a general introduction and~\cite{Schloemer2022} for details regarding the implementation of the mixD setup. After retrieving the maximally entangled state, we employ imaginary time evolution techniques to evolve the state away from $\beta=0$ towards finite temperatures.

Since the infinite temperature states (being projected product states) are of low bond dimension, local approximations of the Hamiltonian (and subsequent exponentiation) will suffer from large projection errors and are of low quality. Hence, we start by employing global methods for a single step in imaginary time, after which the entanglement in the system (and the bond dimension of the thermal MPS) has sufficiently increased to switch to local methods for computational speed. 

In particular, we start with the global Krylov scheme for a single step $\Delta \tau = 0.1$, after which we switch to the local two-site TDVP method~\cite{Paeckel_time}. Convergence of the spin-structure factor in a $40\times 4$ Heisenberg model for various evolution schemes and cutoff parameters is shown in Fig.~\ref{fig:conv}. 
Starting solely via TDVP from the maximally entangled state shows poor convergence towards the global Krylov result (purple curve), even for low weight cutoffs of $w_T = 10^{-10}$ (blue curve). A single Krylov step followed by TDVP steps (denoted here the hybrid scheme) is seen to be sufficient to converge towards the purely global evolution. 

For our mixD calculations, we fix weight cutoffs to $w_K = 10^{-7}$ and $w_T = 10^{-10}$ for the Krylov and TDVP imaginary time evolution, respectively, where convergence is observed. Reconstructions of the spin-Hamiltonian in squeezed space using the $J_1^x - J_1^y - J_2$ Heisenberg model are done with identical truncation parameters. For the $J_1^x - J_1^y - J_2$ Heisenberg model of size $20\times 4$, we choose $w_K = 10^{-6}$ and $w_T = 10^{-9}$ for both the initial simulation as well as the reconstruction. \\ \\

\twocolumngrid
\bibliography{quant_frustr}

\onecolumngrid
\pagebreak
\widetext

\end{document}